\documentclass[11pt,a4paper]{article}
\usepackage{t1enc}
\usepackage[latin1]{inputenc}
\usepackage[english]{babel}
\usepackage{yfonts}
\usepackage{bbm}
\usepackage{bm}
\usepackage{amssymb}
\usepackage{mathrsfs}
\usepackage[all,cmtip,line]{xy}
\usepackage{epstopdf}
\usepackage{graphicx}
\usepackage{hyperref}
\hypersetup{final=true}
\newcommand{\bra}[1]{\langle #1|}
\newcommand{\bbra}[1]{\langle \! \langle #1|}
\newcommand{\ket}[1]{| \, #1\rangle}
\newcommand{\kett}[1]{| \, #1\rangle \! \rangle}

\newcommand{\mf}[1]{\mathfrak{#1}}
\newcommand{\hmf}[1]{\hat{\mathfrak{#1}}}
\newcommand{\be}[0]{\begin{equation}}
\newcommand{\ee}[0]{\end{equation}}
\newcommand{\eqref}[1]{(\ref{#1})}

\newenvironment{definition}[1][Definition]{\begin{trivlist}
\item[\hskip \labelsep {\bfseries #1}]}{\end{trivlist}}

\begin{document}

\title{New WZW D-branes from the algebra of Wilson loop operators}
\author{\small RUNHETC-2009-11 \\ \\ \\Samuel Monnier\footnote{monnier@physics.rutgers.edu} \\ \\ \small New High Energy Theory Center, Department of Physics and Astronomy, \\
\small Rutgers, the State University of New Jersey \\
\small 126 Frelinghuysen Road, Piscataway, NJ 08854, USA \\ \small \& \\
\small Université de Genève, Section de Mathématiques, \\ \small 2-4 Rue du Lièvre, 1211 Genève 24, Switzerland}

\maketitle

\begin{abstract}
We investigate the algebra generated by the topological Wilson loop operators in WZW models. Wilson loops describe the nontrivial fixed points of the boundary renormalization group flows triggered by Kondo perturbations. Their enveloping algebra therefore encodes all the fixed points which can be reached by sequences of Kondo flows. This algebra is easily described in the case of SU(2), but displays a very rich structure for higher rank groups. In the latter case, its action on known D-branes creates a profusion of new and generically non-rational D-branes. We describe their symmetries and the geometry of their worldvolumes. We briefly explain how to extend these results to coset models.
\end{abstract}

\section{Introduction}

The classification of all boundary conditions (or D-branes) compatible with a given two dimensional conformal field theory (CFT) is still an unsolved problem. In view of the numerous applications of D-branes in string theory, as well their relevance to the description of critical condensed matter systems with boundaries, such a classification would be highly desirable. Yet, even a sketch of a program to achieve it is lacking. Among CFT's, rational conformal field theories (RCFT's) are especially simple as they involve only a finite number of primary fields. So far, all the known examples of RCFT's can be constructed from Wess-Zumino-Witten or coset models\footnote{Recent results indicate that other types of RCFT's might exist, see \cite{Dovgard:2008zn}.}. 
The RCFT's for which the D-branes have been completely classified are the minimal models \cite{Cardy:1989ir}, the free boson at the self-dual radius (which is equivalent to the $SU(2)$ WZW model at level 1) \cite{Gaberdiel:2001xm} and the free boson at an irrational radius (as a multiple of the self-dual radius) \cite{Janik:2001hb}. Compared to the full set of WZW and coset models, this list is disappointingly short.

In this paper, we will explore the algebra of topological one-dimensional defects on the worldsheet, a tool that is likely to play a central role in a future classification. We will see that this algebra allows to construct many previously unsuspected D-branes, and that the classification problem for this particular class of D-branes amounts to finding a complete set of generators and relations of the algebra of defects. 
\\

Topological defects can be continuously deformed on the worldsheet without affecting correlation functions. As a result, the set of topological defects supported on cycles in a given homotopy class naturally carries a multiplicative structure. Indeed, two such defects can be deformed so that they lie closer and closer, eventually becoming indistinguishable from a single defect. In the same spirit, topological defects linking a boundary supporting a D-brane can be pushed on it, yielding a new D-brane. Topological defects therefore act on the set of D-branes, and this action is compatible with the multiplication. Practically, this multiplicative structure can be represented on the state space of the CFT by an algebra of defect operators. The action of defect operators on boundary states reproduces the action of defects on D-branes. 

In the case where defects and D-branes are required to preserve a rational vertex subalgebra of the spectrum generating algebra of the theory, a categorical framework allows to classify them completely \cite{Frohlich:2006ch, Fuchs:2002cm}. We will call such defects and D-branes rational. However, the Virasoro algebra associated to the conformal symmetry is generally not a rational vertex subalgebra of the spectrum generating algebra of WZW and coset models. As a result, the generic D-branes compatible with the conformal symmetry are not rational. This simple fact is at the heart of the difficulties in classifying the conformal D-branes, and explain why the categorical framework as it stands is of little use to solve this problem. 

Even if some constructions exist \cite{Blakeley:2007gu}, there is no straightforward way of building directly non-rational D-branes in a rational theory. In contrast, thanks to their multiplicative structure, it may be extremely easy to construct non-rational defects. Indeed, if a RCFT is rational with respect to two distinct vertex algebras, then the product of two rational defects, each preserving one of the vertex algebras, will in general not be a rational defect. In turn, once non-rational defects are known, their action on a rational D-brane will in general yield non-rational D-branes.

As noted above, all the known RCFT's are constructed from affine Kac-Moody algebras, so we will only consider this type of vertex algebras here. We are not aware of a classification of the rational vertex subalgebras of a Kac-Moody algebra\footnote{To simplify the notation, we write $\hmf{g}$ both for the infinite dimensional Lie algebra and the associated vertex algebra. It should be clear from the context which algebra is relevant.} $\hmf{g}$, but there is a natural class of such subalgebras, of the form $\hmf{a} \otimes \hmf{g}/\hmf{a}$, where $\hmf{a}$ is (the vertex algebra associated to) an affine Kac-Moody subalgebra of $\hmf{g}$ and $\hmf{g}/\hmf{a}$ is the coset vertex algebra, built out of all fields in $\hmf{g}$ commuting with $\hmf{a}$. The corresponding rational topological defects also have a straightforward physical interpretation: they are Wilson loops on the worldsheet, namely worldlines of particles whose spin (valued in the horizontal subalgebra $\mf{a}$) couples minimally to the Kac-Moody current. In this paper, we will investigate the algebra generated by all Wilson loops associated to affine Kac-Moody algebras embeddings $\hmf{a} \subset \hmf{g}$. This algebra will be referred to as the Wilson algebra in the sequel.

Of course, we do not expect the Wilson algebra to include all the topological defects of the WZW model. There are in particular some defects associated to $U(1)$ subgroups \cite{Fuchs:2007tx} which will not be included. (These defects should create the D-branes of \cite{Blakeley:2007gu} when acting on maximally symmetric D-branes.) The Wilson algebra is however an interesting truncation of the algebra of topological defects, because the Wilson loops operators encode information about the fixed points of the (possibly symmetry breaking) Kondo boundary renormalization group flows \cite{Bachas:2004sy, Alekseev:2007in}. Indeed, these flows are always of the form 
$$
n\mathbbm{1} \ket{B} \rightarrow W \ket{B} \;, \quad n \in \mathbbm{N} \;,
$$
where $\ket{B}$ is any boundary state, $n\mathbbm{1}$ is the operator corresponding to $n$ copies of the trivial defect, and $W$ is the operator associated to a Wilson loop. As a result, products of Wilson loops describe the fixed points resulting from sequences of Kondo flows and the set of all these fixed points is encoded in the structure of the Wilson algebra. It is especially remarkable that the classification of (a subset of) the conformal fixed points of a theory reduces to a purely algebraic problem. 

Ideally, we would like to find a complete set of generators and relations for the Wilson algebra. With this information, we could classify all the D-branes which can be obtained from a given known D-brane by a sequence of Kondo flows. While a complete set of generators and some relations can be written straightforwardly, finding all the relations is generally a difficult problem that we will be able to solve only for the WZW models based on $SU(2)$ and on $SU(3)$ at level 1. Unfortunately, in these cases, no previously unknown D-brane is generated by the Kondo flows. The investigation of the next examples in term of complexity, namely $SU(3)$ at level 2 and $U\!Sp(4)$ at level 1 will show that new D-branes are generated in these cases, but we do not know whether the set of found relations is complete, impairing a classification of these new D-branes.

In complement to this low-level algebraic investigation, we also study the action of the Wilson algebra at large level, where a geometric interpretation of the D-branes is available. This analysis makes it clear that arbitrary products of Wilson loops generate new D-branes. Indeed, we can associate to each element in the Wilson algebra a submanifold $M$ of the target space Lie group $G$, such that its action on a D-brane with worldvolume $M'$ is given by the pointwise product of $M$ and $M'$. Similarly, the manifold corresponding to the product of two elements in the Wilson algebra is given by the pointwise product of their associated manifolds. Under this correspondence, Wilson loops are mapped to conjugacy classes in the simple subgroup $A$ corresponding to the preserved Kac-Moody subalgebra $\hmf{a} \subset \hmf{g}$. As a result, the action of the Wilson algebra on a D0-brane boundary state produces D-branes which worldvolumes are pointwise products of conjugacy classes of arbitrary simple subgroups of $G$. Rational D-branes of this type were described in \cite{Quella:2002ct, Quella:2002ns}, in the case of a sequence of embeddings $A_1 \subset ... \subset A_k \subset G$. Our results show that when the subgroups do not form a sequence of embeddings, the corresponding D-branes still exist, but are non-rational.

The paper is organized as follows. In section \ref{SecReminder} we review the concept of topological defect. We also briefly present a certain class of defects and D-branes of the WZW models. For reference, the Lie theoretic notation is introduced in section \ref{SecWLOp}. In section \ref{SecWilsAlg} we define the Wilson algebra and derive universal relations. We then study the action of the Wilson algebra on the boundary states of the WZW model in section \ref{SecNWZWB}. We argue that any element of the Wilson algebra produces a consistent boundary state when acting on the D0-brane boundary state and we study the symmetries of these new D-branes. Section \ref{SecGeom} contains the large level geometrical study and section \ref{SecExamples} a few low-level examples, for which it is possible to enumerate the generators of the Wilson algebra, make their action explicit and uncover extra relations. We discuss very briefly in section \ref{SecGener} how these results can be applied to WZW models with non-diagonal modular invariant and coset models.

\section{A reminder about D-branes and defects in WZW models}

\label{SecReminder}

\subsection{Topological defects}

\begin{figure}[!htb]
\centering
\includegraphics[scale=.8]{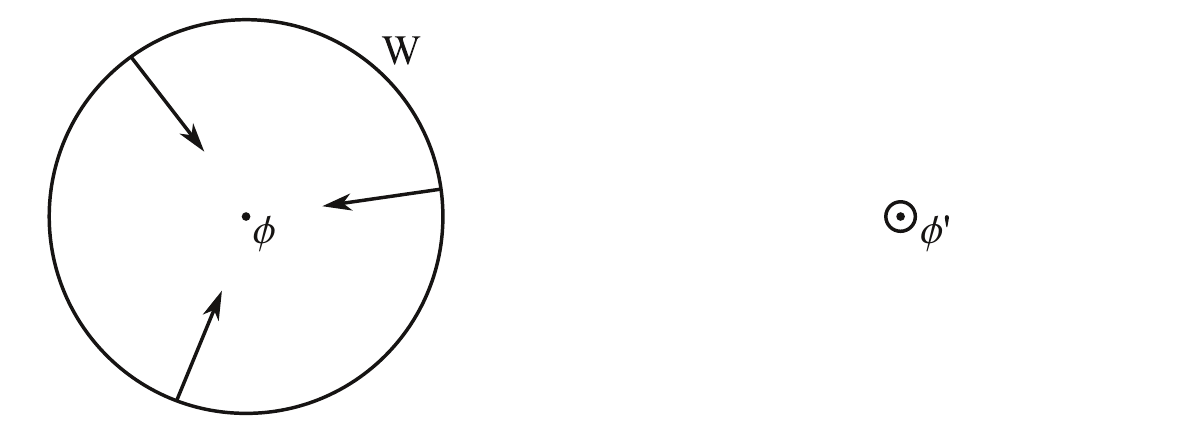}
\caption{Pick a state $v$ in the closed string state space $\mathcal{H}$ of the CFT. By the state-operator correspondence, $v$ gives rise to a bulk field $\phi$. Imagine encircling the insertion of $\phi$ by a topological defect $W$. Because of the topological property, the defect can be shrunk around the insertion, leaving macroscopically a new bulk field $\phi'$ associated to a vector $v' \in \mathcal{H}$. We define the action of the defect on $\mathcal{H}$ by $W(v) = v'$.}
\label{Fig1}
\end{figure}

\begin{figure}[!htb]
\centering
\includegraphics[scale=.8]{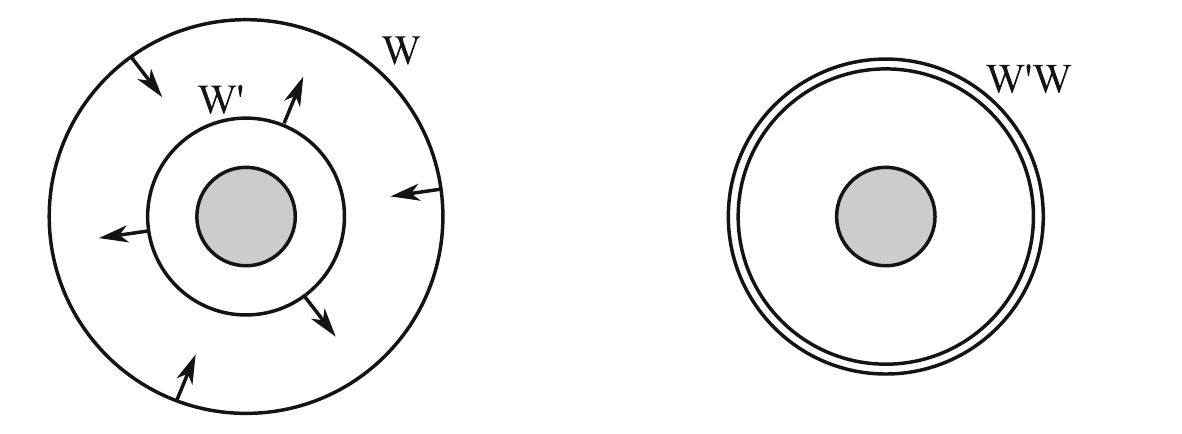}
\caption{Similarly, two topological defects $W$ and $W'$ supported on cycles in the same homotopy class can be deformed until they form macroscopically a single defect $W'W$, providing the set of topological defects with a multiplicative law. Using the topological property, it is straightforward to check that this multiplicative law translates into the composition of the corresponding operators.}
\label{Fig2}
\end{figure}

\begin{figure}[!htb]
\centering
\includegraphics[scale=.8]{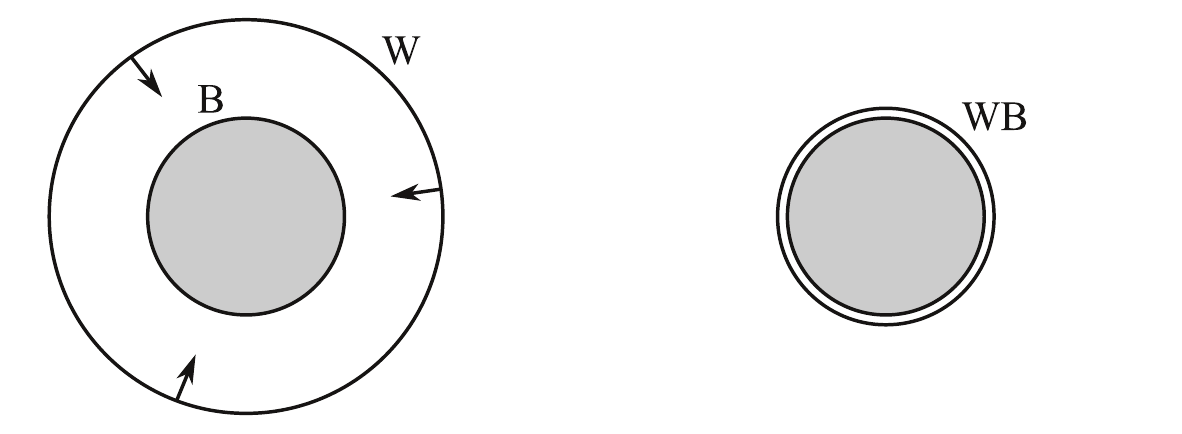}
\caption{Finally, a topological defect $W$ encircling a boundary supporting a D-brane $B$ can be shrunk onto the boundary, yielding a new D-brane $WB$. The action of the defect on the D-brane is given by the action of the defect operator on the corresponding boundary state, seen as an element of (a completion of) $\mathcal{H}$.}
\label{Fig3}
\end{figure}

A defect is called topological if the correlation functions involving it depend only on the homotopy class of the cycle supporting it. The set of topological defects carries a multiplicative structure (figure \ref{Fig2}), and can be mapped injectively\footnote{We are touching here a subtle point. Strictly speaking this statement, which holds for rational defects \cite{Frohlich:2006ch}, is not true for non-rational defects. There is at least one example of a one-parameter family of distinct defects which correspond to the same operator \cite{Fuchs:2007tx}. However this example is somewhat pathological, as the family is constructed from a non-hermitian perturbation of a single defect. We will not encounter any defect of this type: our non-rational defects will always be expressed as sums and products of rational defects. It seems therefore reasonable to consider that the map from the Wilson algebra to operators on $\mathcal{H}$ is injective, and we will always interpret an equality of defect operators as an equality of defects. But we cannot completely exclude the possibility that some of the non-rational defects operators and boundary states found in this paper actually correspond to several distinct defects and D-branes.} into the algebra of bounded operators on the closed string state space $\mathcal{H}$ (figure \ref{Fig1}). Furthermore, topological defects have a natural action on D-branes (figure \ref{Fig3}). These ideas are developed rigorously in \cite{Frohlich:2006ch}.

Defects can be added, just like D-branes can be stacked, and this operation corresponds to the addition of the corresponding operators. They also have to satisfy positivity and integrality conditions related to the Cardy condition for boundary states when they are inserted in a cylinder amplitude. Therefore, the difference of two defects is in general not a defect. As a result, the set of defects forms an algebra over the semiring\footnote{Recall that a semiring satisfies the same axioms as a ring, except that addition is not required to be invertible. Algebras and modules can be defined over semirings in a straightforward way, see the appendix \ref{SecSemi}.} $\mathbb{N}$.

\subsection{Wilson loop operators}

\label{SecWLOp}

Let us now specialize the discussion to a WZW model based on a compact, simple and simply-connected Lie group $G$ and fix some notation. 

Denote by $\mf{g}$ the Lie algebra of $G$ and $\hmf{g}$ the corresponding untwisted affine Kac-Moody algebra. We will write $\hmf{g}_k$ when we are specializing the level of $\hmf{g}$ to the positive integer value $k$. Let $\mf{a}$ be a simple subalgebra of $\mf{g}$. $\{e_a\}$, $a = 1,...,{\rm dim}(\mf{g})$ and $\{\tilde{e}_i\}$, $i = 1,...,{\rm dim}(\mf{a})$ will denote bases of $\mf{g}$ and $\mf{a}$ orthonormal with respect to their respective Killing forms, and $\pi_a^{\;i}$ the matrix elements of the projection of $\mf{g}$ onto $\mf{a}$.

The embedding $\mf{a} \subset \mf{g}$ generates an embedding of affine Kac-Moody algebras $\hmf{a}_{xk} \subset \hmf{g}_k$, where $x$ is the embedding index of $\mf{a}$ (see the section 13.7 of \cite{CFT1997}). We will denote the generators of $\hmf{g}_k$ by $J^a_n$ with $n \in \mathbbm{Z}$, and write $J^a(z) = \sum_{n \in \mathbbm{Z}} J^a_n z^{-1-n}$. The generators of $\hmf{a}_{xk}$ are $\tilde{J}^i_n = J^a_n \pi_a^{\;i}$. The spectrum generating algebra of the WZW model is composed of two copies of $\hmf{g}_k$. The generators of the second (antiholomorphic) copy will be written $\bar{J}^a_n$.

Let $P^\mf{a}$ be the set of dominant weights of $\mf{a}$, $\rho_\sigma$ be the representation of $\mf{a}$ of highest weight $\sigma \in P^\mf{a}$ and $V^{\mf{a}}_\sigma$ the corresponding irreducible module. Similarly, let $P^{\hmf{a}_{xk}}$ be the set of integrable weights of $\hmf{a}$ at level $xk$ and $H^{\hmf{a}_{xk}}_\sigma$ the integrable module with highest weight $\sigma \in P^{\hmf{a}_{xk}}$. As we will always work at a fixed level, we will identify the affine weights of $\hmf{a}_{xk}$ with their horizontal part in $P^\mf{a}$. 

Consider the classical WZW model defined on a Riemann surface $\Sigma$ with a holomorphic coordinate $z$. The classical current is defined from the sigma-model map $g : \Sigma \rightarrow G$ by $j^a(z) e_a = -k(\partial_z g)g^{-1}$. We can define a $\mf{a}$-valued current $\tilde{j}^i(z) = j^a(z) \pi_a^{\;i}$. Now let us choose $\sigma \in P^\mf{a}$. The classical Wilson loop associated to $\mf{a}$ and $\sigma$ is the non-local observable given by
\be
\label{ClassWLoop}
w^{\mf{a}}_\sigma = {\rm Tr}_{V_\sigma} {\rm P} \exp \frac{i}{xk} \oint_C dz \tilde{j}^i(z) \rho_\sigma(\tilde{e}_i) \;,
\ee
where $C$ is an oriented loop on $\Sigma$ and P is the path ordering along $C$. This observable is topological, in the sense that it depends on $C$ only through its homotopy class. When $\mf{a} = \mf{g}$, it measures the holonomy of the pull-back $j(z)$ of the Maurer-Cartan connection. In general, it computes the holonomy of the projected connection $\tilde{j}(z)$. As the holonomy is a gauge invariant observable, $w^{\mf{a}}_\sigma$ has vanishing Poisson bracket with the functions $\tilde{j}^i(z)$.

Upon quantization, the closed string states of the WZW model with charge conjugation modular invariant form a Hilbert space isomorphic to
\be
\label{StSpChCModIn}
\mathcal{H} = \bigoplus_{\mu \in P^{\hmf{g}_k}} H^{\hmf{g}_k}_\mu \otimes H^{\hmf{g}_k}_{\mu^\ast} \;.
\ee
The holomorphic current $J(z)$ acts only on the first factors of the summands of $\mathcal{H}$. The quantum counterpart of \eqref{ClassWLoop} is expected to be an operator on $\mathcal{H}$ expressible as a series in $\tilde{J}^i_n$. We also expect the quantization procedure to preserve the classical symmetries of the Wilson loop, so the defect operator should centralize $\hmf{a}_{xk}$. It turns out that these properties allow to determine the quantum defect operator uniquely \cite{Alekseev:2007in} (see also \cite{Bachas:2004sy} for a perturbative approach to the quantization of a one parameter deformation of these loop operators). The defect operator $W^{\mf{a}}_\sigma$ is characterized by its eigenvalues on $\hmf{a}_{xk}$-modules:
\be
\label{ActWLoop}
W^{\mf{a}}_\sigma = \chi_\sigma \left ( \frac{-2\pi i}{xk + \check{h}} (\tau + \rho) \right ) \mathbbm{1} = \frac{S_{\sigma\tau}}{S_{0\tau}} \mathbbm{1} \quad {\rm on} \; H^{\hmf{a}_{xk}}_{\tau} \;,
\ee 
where $\check{h}$ and $\rho$ are respectively the dual Coxeter number and the Weyl vector (half the sum of the positive roots) of $\mf{a}$. $\chi_\sigma$ is the character of the representation $\rho_\sigma$, seen as a function on the weight space of $\mf{a}$. In the second equality, we assumed that $\sigma \in P^{\hmf{a}_{xk}}$ and we used a well-known identity (see \cite{CFT1997}, section 14.6.3) to reexpress the character in term of the modular $S$ matrix of $\hmf{a}_{xk}$.

Through an iterative procedure, $W^{\mf{a}}_\sigma$ can indeed be constructed as a series in $\tilde{J}^i(z)$ commuting with $\hmf{a}_{xk}$. This important fact allows to deduce the action of $W^{\mf{a}}_\sigma$ on $\mathcal{H}$ from its action on the integrable highest weight modules of $\hmf{a}_{xk}$. To this end, we should decompose $\mathcal{H}$ as follows:
\be
\label{DecompStSpace}
\mathcal{H} = \bigoplus_{\mu \in P^{\hmf{g}_k}}  \bigoplus_{\sigma \in P^{\hmf{a}_{xk}}} H^{\hmf{a}_{xk}}_\sigma \otimes H^{\hmf{g}/\hmf{a}}_{[\sigma,\mu]} \otimes H^{\hmf{g}_k}_{\mu^\ast} \;,
\ee
where $H^{\hmf{g}/\hmf{a}}_{[\sigma,\mu]}$ are the multiplicity spaces occurring in the decomposition of $\hmf{g}_k$ modules into $\hmf{a}_{xk}$ modules. (These multiplicity spaces are the basic building blocks of the state space of the $\hmf{g}_k/\hmf{a}_{xk}$ coset model.) In this decomposition, the action of $W^{\mf{a}}_\sigma$ is straightforward: it acts by the scalar multiplication \eqref{ActWLoop} on the first factor of each summand in \eqref{DecompStSpace} and trivially on the other factors. Remark that the Wilson loop labeled by the zero weight always acts like the identity operator.

For fixed $\mf{a}$, the algebra generated by the Wilson loop operators is commutative: using \eqref{ActWLoop} and the Verlinde formula, we get
\be
\label{AlgWLoop}
W^{\mf{a}}_\sigma W^{\mf{a}}_\tau = \mathcal{N}^{\mf{a}\;\;\upsilon}_{\sigma\tau} W^{\mf{a}}_\upsilon \;,
\ee
where $\mathcal{N}^{\mf{a}\;\;\upsilon}_{\sigma\tau}$ are the fusion coefficients of $\hmf{a}_{xk}$. Moreover, $W^{\mf{a}}_\sigma$ commutes with the Virasoro vertex subalgebra of $\hmf{g}_k$. Indeed, it is well-known (for instance \cite{CFT1997}, section 18.1) that 
$$
L_n = L^{\hmf{a}_{xk}}_n + L^{\rm coset}_n \;,
$$ 
where $L_n$ and $L^{\hmf{a}_{xk}}_n$ are the Virasoro generators obtained from the Sugawara construction on $\hmf{g}_k$ and $\hmf{a}_{xk}$ respectively, and $L^{\rm coset}_n$ commutes with $\hmf{a}_{xk}$. From the central property of $W^{\mf{a}}_\sigma$, it commutes with $L^{\hmf{a}_{xk}}_n$. From the fact that it is a series in the current of $\hmf{a}_{xk}$, it commutes with $L^{\rm coset}_n$. Therefore it commutes with $L_n$. By the same reasoning, we see that $W^{\mf{a}}_\sigma$ actually commutes with the vertex algebra $\hmf{a}_{xk} \otimes \hmf{g}_k/\hmf{a}_{xk}$, so $W^{\mf{a}}_\sigma$ is a rational defect.

\subsection{Branes in WZW models}

For our purpose, it will be sufficient to consider D-branes carrying D0-brane charge only. The known D-branes of this type can be described in a simple way with the help of Wilson loops.

The maximally symmetric D-branes preserve the diagonal subalgebra of the spectrum generating algebra $\hmf{g}_k \otimes \hmf{g}_k$. This condition translates into \cite{Zuber:2000ia, Gaberdiel:2002my}
$$
(J^a_n + \bar{J}^a_{-n}) \ket{B} = 0 \;.
$$
This equation admits a unique solution up to scalar multiplication in each of the sectors $H_\mu \otimes H_{\mu^\ast}$ of $\mathcal{H}$: the Ishibashi state $\kett{\mu}$. These ``states'' are not normalizable and actually belong to a completion of $H_\mu \otimes H_{\mu^\ast}$. It is however possible to remove the scalar ambiguity in their definition by requiring that 
\be
\label{NormIshi}
\bbra{\mu} q^{\frac{1}{2}(L_0 + \bar{L}_0 - \frac{c}{12})} \kett{\nu} = S_{0\mu}\delta_{\mu\nu} \chi^{\hmf{g}_k}_\mu(q) \;,
\ee
where $\chi^{\hmf{g}_k}_\mu(q)$ is the affine character of $H_{\mu}$, $c$ the central charge of the model and $S_{\mu\nu}$ the elements of its modular S matrix. This convention differs slightly from the usual one found in the literature, which does not include the factor $S_{0\mu}$ on the right hand side. 

The modularity constraint (Cardy's condition) allows to determine which linear combinations of Ishibashi states yield consistent boundary states. The simplest boundary state corresponds to the D0-brane and is given by 
$$
\ket{B_0} = \sum_{\mu \in P^{\hmf{g}_k}} \kett{\mu} \;.
$$
The other maximally symmetric boundary states are recovered by the action of the maximally symmetric Wilson loop operators $W^{\mf{g}}_\nu$ on $\ket{B_0}$:
$$
\ket{B_\nu} = W^{\mf{g}}_\nu \ket{B_0} = \sum_{\mu \in P^{\hmf{g}_k}} \frac{S_{\nu\mu}}{S_{0\mu}} \kett{\mu} \;.
$$
In \cite{Quella:2002ct,Quella:2002ns}, symmetry breaking rational D-branes were constructed. The corresponding boundary states can again be nicely described with Wilson loop operators as follows. Consider a chain of embeddings of semi-simple Lie groups labeled by upper indices: $A^1 \subset A^2 \subset ... \subset A^n \subset G$, with the corresponding embedding chain of Lie algebras: $\mf{a}^1 \subset \mf{a}^2 \subset ... \subset \mf{a}^n \subset \mf{g}$ and of Kac-Moody algebras:  $\hmf{a}_{k_1}^1 \subset \hmf{a}_{k_2}^2 \subset ... \subset \hmf{a}_{k_n}^n \subset \hmf{g}_k$. To a sequence of integrable weights $\sigma_i \in P^{\hmf{a}_{k_i}^i}$ and $\mu \in P^{\hmf{g}_k}$, associate the boundary state\footnote{The construction of \cite{Quella:2002ct,Quella:2002ns} is more general and allows for D-branes twisted by nontrivial automorphisms of $\mf{a}^i$ and $\mf{g}$, but we will not consider them here.} 
\be
\label{CosBoundSt}
\ket{B;\sigma_1,\sigma_2,...,\sigma_n,\mu} = W^{\mf{a}^1}_{\sigma_1}W^{\mf{a}^2}_{\sigma_2}...W^{\mf{a}^n}_{\sigma_n} W^{\mf{g}}_\mu \ket{B_0} \;.
\ee
These boundary states satisfy Cardy's condition \cite{Quella:2002ct,Quella:2002ns}. They preserve the rational vertex subalgebra
$$
\hmf{a}_{k_1}^1 \otimes  \hmf{a}_{k_2}^2/\hmf{a}_{k_1}^1 \otimes ... \otimes \hmf{g}_k/\hmf{a}_{k_n}^n \subset \hmf{g}_k \;.
$$
We will see that they are actually part of a much larger family of generically non-rational D-branes.

\subsection{Wilson operators and Kondo perturbations}

\label{SecWOKondo}

When its supporting cycle $C$ coincides with a boundary component of the worldsheet, the Wilson loop \eqref{ClassWLoop} can be seen as the transfer matrix of the Kondo boundary perturbation \cite{affleck-1995-26}. The Kondo perturbation is marginally relevant and therefore triggers a boundary renormalization group flow. The beautiful idea put forward in \cite{Bachas:2004sy} (and which can be traced back to \cite{Bazhanov:1994ft}) is that this boundary flow is induced from a ``universal'' renormalization group flow acting on defects. Indeed, the defect $d_\sigma \mathbbm{1}$ (consisting of a stack of $d_\sigma = {\rm dim}(V_\sigma)$ copies of the trivial defect) admits a perturbation of the Kondo type. The perturbation defines a family of generically non-conformal defects which can therefore undergo a renormalization group flow. \cite{Bachas:2004sy} showed that, at least in the large $k$ limit, the IR fixed point of this flow is the Wilson loop defect described by the operator $W^{\mf{a}}_\sigma$, and this fact was checked by an explicit integration of the Kondo perturbation at a critical value of the coupling in \cite{Alekseev:2007in}. When the defect is pushed onto a boundary supporting a boundary state $\ket{B}$, the defect flow induces a boundary renormalization group flow
\be
\label{FlowWLoop}
d_\sigma \mathbbm{1}\ket{B} \rightarrow W^{\mf{a}}_\sigma \ket{B} \;.
\ee
Such defect flows are universal, because they induce boundary renormalization group flows when acting on \emph{any} boundary state $\ket{B}$. They provide a justification of the ``absorption of boundary spin'' principle of Affleck-Ludwig \cite{Affleck:1990by}, an empirical rule used to determine the fixed points of the Kondo flow. Its generalization to coset models \cite{Fredenhagen:2002qn, Fredenhagen:2003xf} can be understood in term of defect operators in a similar way \cite{BM}. \eqref{FlowWLoop} also shows that all the boundary states \eqref{CosBoundSt} are fixed points of sequences of boundary renormalization group flows starting from a stack of D0-branes (see \cite{Monnier:2005jt} for a different point of view on this fact). \\

There is another set of universal boundary perturbations of the WZW model, namely the exactly marginal perturbations corresponding to translations on the group manifold. Such a translation is implemented on the boundary states by the exponential of the action of the horizontal Lie algebra of $\hmf{g}_k$:
$$
\ket{B} \rightarrow \exp(\lambda_a J^a_0)\ket{B} \;, \quad g = \exp(\lambda_a e^a) \in G \;.
$$
Geometrically, $t_g = \exp(\lambda_a J^a_0)$ translates the worldvolume of the D-brane by the left action of $g$ on the group manifold. 

Despite of our notational distinction, it should be emphasized that the operator $t_g$ is nothing but a Wilson operator associated to the commutative subalgebra of $\mf{g}$ generated by $\lambda_a e^a$. Indeed, when $\mf{a}$ is commutative, the path ordering in \eqref{ClassWLoop} has no effect. The Wilson loop is easily quantized by replacing the classical current $j$ by its quantum counterpart $J$ and performing the integration explicitly. The quantum operator obtained is exactly $t_g$. It will therefore be natural to include the operators $t_g$ in our study of the algebra of Wilson operators. The corresponding defect will be called a ``group defect'' in the sequel. \\

\section{The Wilson algebra}

\label{SecWilsAlg}

\subsection{Definition}

\label{SemiDef}

It is natural to study the algebra generated by the defect operators described in the previous section. Denote by $\mathcal{W}$ the algebra over the semiring $\mathbb{N}$ generated by 
\begin{itemize}
	\item the Wilson operators $W^{\mf{a}}_\sigma$, for all simple subalgebras $\mf{a} \subset \mf{g}$ and all non zero integrable weights $\sigma$ of $\hat{\mf{a}}_{xk}$,
	\item the groups defect operators $t_g = \exp(\lambda_a J^a_0)$ for $g = \exp(\lambda_a e^a) \in G$.
\end{itemize}

The Wilson operators associated to direct sums of Lie algebras are products of Wilson operators associated to the irreducible summands. We also saw that the Wilson operators associated to commutative subalgebras of $\hmf{g}$ are the group defects $t_g$. Hence the Wilson algebra as defined above includes all the Wilson operators associated to reductive subalgebras of $\mf{g}$. 

Remark that all the generators are expressed as series in the current $J$ which have well-defined actions on the integrable highest weight modules of $\hmf{g}$. The induced action of $\mathcal{W}$ is chiral: $\mathcal{W}$ acts nontrivially only on the first factors of the components $H^{\hmf{g}_k}_\mu \otimes H^{\hmf{g}_k}_{\mu^\ast}$ of $\mathcal{H}$. Note also that while each of the families $\{W^{\mf{a}}_\sigma\}$ for a fixed $\mf{a}$ generates a commutative algebra, $\mathcal{W}$ is not commutative.

\subsection{Relations}

\label{WRRel}

We can immediately write down some relations for the algebra $\mathcal{W}$. First, obviously, 
$$
t_g t_{g'} = t_{gg'} \;.
$$
For a fixed sublgebra $\mf{a}$, the relations \eqref{AlgWLoop} hold:
$$
W^{\mf{a}}_\sigma W^{\mf{a}}_\tau = \mathcal{N}^{\mf{a}\;\;\upsilon}_{\sigma\tau} W^{\mf{a}}_\upsilon \;.
$$
If $g\mf{a}g^{-1}$ is the subalgebra obtained from $\mathfrak{a}$ by the adjoint action of $g$ on $\mf{g}$, then 
$$
W^{g\mf{a}g^{-1}}_\sigma = t_g W^{\mf{a}}_\sigma t_{g^{-1}} \;,
$$
as is obvious from the representation of $W^{\mf{a}}_\sigma$ as a series in the current of $\hat{\mf{a}}_{xk}$. Therefore we need as generators only the Wilson operators associated to one element in each conjugacy class of simple subalgebras of $\mf{g}$. After reducing the number of generators, we should still include relations of the type $W^{\mf{a}}_\sigma = t_g W^{\mf{a}}_\sigma t_{g^{-1}}$ whenever $g \in N(A)$, the normalizer of $A$. 

If $\mf{a}$ and $\mf{b}$ are two commuting subalgebras, then 
$$
W^{\mf{a}}_\sigma W^{\mf{b}}_\nu = W^{\mf{b}}_\nu W^{\mf{a}}_\sigma \;.
$$
This again follows from the representations of $W^{\mf{a}}_\sigma$ and $W^{\mf{b}}_\nu$ as series in the commuting currents $J|_{\hat{\mf{a}}}$ and $J|_{\hat{\mf{b}}}$, respectively. 
\\

We will describe now a less obvious type of relations. It is well known (see for instance section 14.6.4 of \cite{CFT1997}) that the outer automorphism group $\mathcal{O}(\hat{\mf{a}}_\kappa)$ of an affine Kac-Moody algebra $\hat{\mf{a}}$ at level $\kappa$ is isomorphic to the center $Z_A$ of $A$. Indeed, let $\hat{\Omega} \in \mathcal{O}(\hat{\mf{a}}_\kappa)$ and denote the induced operator on the weight space of $\mf{a}$ by $\Omega$. The affine fundamental weight $\hat{\omega}_0$ projects onto the zero weight of $\mf{a}$, and 
$$
\exp \left ( -2\pi i \sum_{i=1}^{r} \Omega(0)^i \tilde{e}_i \right ) = z_\Omega \in Z_A \;,
$$
where we assumed that the first $r = {\rm rank}(\mf{a})$ elements in the orthonormal base $\{\tilde{e}_i\}$ of $\mf{a}$ generate the Cartan subalgebra. The modular $S$ matrix of $\hat{\mf{a}}_\kappa$ satisfies 
$$
S_{\Omega(\sigma)\tau} = S_{\sigma\tau} \exp(-2\pi i (\Omega(0),\tau)) \;,
$$
where $\sigma,\tau$ are integrable weights at level $\kappa$ and $(.,.)$ is the bilinear form induced by the Killing form of $\mf{a}$.

Now recall that the eigenvalues of the Wilson operator $W^{\mf{a}}_\sigma$ are given by
$$
W^{\mf{a}}_\sigma = \frac{S_{\sigma\tau}}{S_{0\tau}} \mathbbm{1} \qquad {\rm on} \; H^{\hmf{a}_\kappa}_\tau \;.
$$
We immediately deduce a new set of relations in the Wilson algebra: 
$$
W^{\mf{a}}_{\Omega(\sigma)} = t_{z_\Omega} W^{\mf{a}}_\sigma \;.
$$
In particular, the Wilson operators $W^{\mf{a}}_{\Omega(0)}$ lying in the orbit of the zero weight coincide with $t_{z_\Omega}$. The associated topological defects are ``group-like'' defects in the terminology of \cite{Frohlich:2006ch}.

Because of these new relations, we can further reduce the set of generators of $\mathcal{W}$, by including only Wilson operators labeled by a chosen representative in each orbit of the outer automorphism group of $\hmf{a}$. Let us summarize here the set of generators and relations so far.
\be
\label{GenRel}
\renewcommand{\arraystretch}{1.8}
\begin{array}{l|l}
 {\rm Generators} & {\rm Relations} \\
\hline
W^{\mf{a}}_\sigma \;, \;\; t_g & t_g t_{g'} = t_{gg'} \\
& W^{\mf{a}}_\sigma W^{\mf{a}}_\tau = \mathcal{N}^{\mf{a}\;\;\upsilon}_{\sigma\tau} W^{\mf{a}}_\upsilon \\
& W^{\mf{a}}_\sigma = t_g W^{\mf{a}}_\sigma t_{g^{-1}} \;\; {\rm for} \; g \in N(A) \\
& W^{\mf{a}}_\sigma W^{\mf{b}}_\nu = W^{\mf{b}}_\nu W^{\mf{a}}_\sigma \;\; {\rm if} \; [\mf{a},\mf{b}] = 0 \\ 
\end{array}
\ee
with $g,g' \in G$, $\mf{a}$ in a set of representatives of each conjugacy class of simple subalgebras of $\mf{g}$, $\sigma, \tau$ in a set of representatives of each orbit of the outer automorphism group of $\hmf{a}$, except the orbit of the zero weight.

At least in some cases, further relations hold (see for instance the solved case of $SU(3)$ at level one in section \ref{su3lev1}), but we see no way of deriving them systematically.

\section{New WZW branes}

\label{SecNWZWB}

In the previous section, we defined the Wilson algebra as the enveloping algebra of all the Wilson defect operators associated reductive subalgebras of $\mf{g}$. In the rational case, the set of rational D-branes forms a module for the algebra of rational defects. A natural conjecture is therefore that the set all D-branes of the WZW model should carry an action of algebra of all WZW defects, and in particular an action of the Wilson algebra. Put differently, given any boundary state $\ket{B}$, each element in the induced module $\mathcal{W}\ket{B}$ should be a consistent boundary state. For definiteness, we will consider only the case when $B$ is the D0-brane $B_0$. 

A very special case is $G = SU(2)$. $\mf{su}(2)$ does not contain any proper simple subalgebra, so the generators of the Wilson algebra are the maximally symmetric Wilson operators $W^{ms}_\mu$ indexed by integrable weights of $\hmf{su}_k(2)$, as well as the group defect operators $t_g$, $g \in SU(2)$. The maximally symmetric Wilson operators commute with the action of the Kac-Moody algebra, hence with $t_g$ for any $g \in SU(2)$. We can therefore write a typical monomial of the Wilson algebra as $nt_g W^{ms}_\mu$ with $n \in \mathbb{N}$ and $g\in SU(2)$. Upon acting on the D0-brane state $\ket{B_0}$, this element produces a stack of $n$ maximally symmetric D-branes of label $\mu$, translated on the group manifold by the left action of $g$. Hence the structure of the Wilson algebra can be completely elucidated in this case and does not lead to the discovery of new D-branes.

For higher rank groups, most of the elements in $\mathcal{W}\ket{B_0}$ do not correspond to any previously known boundary state. Therefore we deduce the existence of an abundance of previously unsuspected D-branes.

To prove the consistency of these new D-branes, we would have to show that the boundary states $\mathcal{W}\ket{B_0}$ satisfy the sewing conditions of \cite{Lewellen:1991tb}, in particular the Cardy condition \cite{Cardy:1989ir} for cylinder amplitudes. A direct verification seems very difficult. We will give some physical arguments for the consistency of these boundary states later in this section.

\subsection{Symmetries}

\label{SecSym}

Let us first look at the symmetries of the conjectured D-branes. A D-brane preserves a vertex subalgebra $\hmf{v}$ if the corresponding boundary state $\ket{B}$ satisfies gluing conditions of the form
\be
\label{GlueCondGen}
(I(X_{n}) - (-1)^{h(X)}\bar{I}(X_{-n}))\ket{B} = 0 \;.
\ee
In the previous equation, $X$ is a field of $\hmf{v}$ of conformal dimension $h(X)$, and $X_{n}$ are its Laurent modes. $I$ is an embedding realizing $\hmf{v}$ as a vertex subalgebra of the holomorphic copy of $\hmf{g}_k$ of the spectrum generating algebra of the WZW model. Similarly, $\bar{I}$ is an embedding of $\hmf{v}$ into the antiholomorphic copy of $\hmf{g}_k$. 

The D0-brane preserves $\hmf{g}_k$, with the trivial embeddings $I = \bar{I} = \mathbbm{1}_{\hmf{g}_k}$. Moreover, recall that $\mathcal{W}$ acts only on the holomorphic sector of the theory, so $\mathcal{W}$ commutes with $\bar{I}(X_{-n})$ for any $\bar{I}$ and $X_{-n}$. Let us consider only the ``untwisted'' case, for which $I = \bar{I}$. Then a necessary and sufficient condition for $\hmf{v}$ to be preserved by a D-brane $WB_0$, $W \in \mathcal{W}$, is that $W$ centralizes $I(\hmf{v})$. 

We have seen in section \ref{SecWLOp} that the Wilson loop operators commute with the Virasoro subalgebra of $\hmf{g}_k$. Moreover, the Sugawara construction implies that the zero modes of the current commutes with the Virasoro algebra as well, so we have $[t_g,L_n] = 0$. Hence the Virasoro subalgebra is preserved: all the D-branes in the set $\mathcal{W}B_0$ are conformal. (This is equally true for any D-brane which boundary state is obtained from the action of an element of $\mathcal{W}$ on a conformal boundary state.)

Does a typical D-brane have more symmetries ? Let $\hmf{v}_{\rm sym}$ be the maximal vertex subalgebra of $\hmf{g}_k$ preserved by all the D-branes in $\mathcal{W}B_0$. The requirement that $\hmf{v}_{\rm sym}$ commutes with the action of $t_g$, $g \in G$ implies that it is a vertex subalgebra of the Casimir algebra of $\hmf{g}_k$ (see the part 7 of \cite{Bouwknegt:1992wg}). The latter is generated by the fields 
$$
C_n(z) = C^{a_1...a_n} :J^{a_1}...J^{a_n}:(z) \;,
$$
where $:...:$ is the normal ordering and $C^{a_1...a_n}$ are the components of the order $n$ Casimir operator in the enveloping algebra $U(\mf{g})$. However, except for $C_2(z) = L(z)$, the higher Casimirs do not belong to $\hmf{a}_{xk} \times \hmf{g}_k/\hmf{a}_{xk}$, so they do not commute with the Wilson operators.

To see this, remark that by definition, the first order pole of the operator product expansion (OPE) of the current $J^i(z)$ of $\hmf{a}_{xk}$ with any element of $\hmf{a}_{xk} \times \hmf{g}_k/\hmf{a}_{xk}$ involves only fields of $\hmf{a}_{xk}$. The OPE of $J^i(z)$ with $C_n(w)$ is given by \cite{Bais:1987dc}
\be
\label{OPECurCas}
J^i(z)C_n(w) \simeq \frac{{\rm cst}}{z-w}C^{ia_1...a_{n-1}}:J^{a_1}...J^{a_{n-1}}:(w) + {\rm reg} \;,
\ee
where the constant depends on $k$. The first term of \eqref{OPECurCas} clearly does not belong to $\hmf{a}_{xk}$.

We deduce that the D-branes created from the action of a typical element of $\mathcal{W}$ on a D0-brane preserve only the Virasoro vertex subalgebra of $\hmf{g}_k$. They break as much symmetry as is consistent with conformal symmetry and are non-rational.

Now we would like to give some evidence for the existence these new D-branes.

\subsection{Construction via boundary RG flows}

All the D-branes in the set $\mathcal{W}B_0$ can be constructed by sequences of boundary renormalization group flows of the Kondo type. We saw in section \ref{SecWOKondo} that the Kondo flows are universal boundary renormalization group flows of the form 
\be
\label{KonFlowBSt}
d_\sigma \ket{B} \rightarrow W^{\mf{a}}_\sigma \ket{B} \;,
\ee
sending a stack of $d_\sigma = {\rm dim}(V_\sigma)$ D-branes of type $B$ on the D-brane which boundary state is given by $W^{\mf{a}}_\sigma \ket{B}$. The crucial point is that this flow occurs between topological defects: $d_\sigma \mathbbm{1} \rightarrow W^{\mf{a}}_\sigma$, so \eqref{KonFlowBSt} is universally valid for \emph{any} boundary state $\ket{B}$ \cite{Bachas:2004sy}.

Consider now a sequence of Wilson operators $W_1$, $W_2$, ...,$W_k$, associated to representations of simple subalgebras of $\mf{g}$ of dimensions $d_1$, $d_2$, ...,$d_k$. From the universality of \eqref{KonFlowBSt}, we immediately see that we have a sequence of flows
$$
d_1d_2...d_k\ket{B_0} \rightarrow d_1d_2...d_{k-1}W_k\ket{B_0} \rightarrow ... \rightarrow d_1W_2...W_k\ket{B_0} \rightarrow W_1W_2...W_k\ket{B_0} \;,
$$
so that a generic element of $\mathcal{W}\ket{B_0}$ can be obtained from a stack of $d_kd_{k-1}...d_1$ D0-branes. While this argument is a strong evidence for the existence of these boundary states, it is not a rigorous proof, because the validity of \eqref{KonFlowBSt} for any $\ket{B}$ rests on physical arguments. 

K-theory predicts that D-branes on a Lie group or rank $r$ can carry $2^{r-1}$ different types of charges (one of which is the D0-brane charge), but so far D-branes carrying all of the possible charges are known only for $SU(2)$ and $SU(3)$ (see \cite{Monnier:2008jj} for a recent account). One may wonder if the D-branes constructed here may carry some of the missing charges. Unfortunately, the sequence of flows displayed above shows that they cannot carry any charge other than the D0-brane charge. More generally, the D-branes obtained from the action of the Wilson algebra on a D-brane carrying a certain charge can only carry multiples of this charge.

\subsection{Identities between partition functions}

The topological nature of the Wilson loops can be used to derive identities between cylinder amplitudes and open string partition functions. Consider the following cylinder amplitude, for $W_1$, $W_2$ and $W_3$ in $\mathcal{W}$: 
$$
\bra{B_0}W_1^\dagger q^{\frac{1}{2}(L_0 + \bar{L}_0 - \frac{c}{12})} W_2W_3\ket{B_0} = \bra{B_0}W_1^\dagger W_2 q^{\frac{1}{2}(L_0 + \bar{L}_0 - \frac{c}{12})} W_3\ket{B_0} \;, \quad q = \exp 2\pi i \tau \;,
$$
$\tau$ being the modular parameter of the cylinder. The equality above stems from the fact that any element of $\mathcal{W}$ commutes with the Virasoro vertex subalgebra of $\hmf{g}_k$ (see section \ref{SecSym}). After performing a modular transformation, we deduce that the partition function of open strings stretching between the D-branes $W_2^\dagger W_1 B_0$ and $W_3 B_0$ coincides with the partition function for open strings between $W_1B_0$ and $W_2W_3B_0$.

This property can be used to give another piece of evidence for the existence of the conjectured D-branes. Consider two boundary states 
$$
W^{\mf{a}}_\upsilon W^{\mf{a}}_\sigma W^{\mf{g}}_\mu \ket{B_0} = \mathcal{N}_{\sigma \upsilon}^{\hmf{a}_k \; \tau} W^{\mf{a}}_\tau W^{\mf{g}}_\mu \ket{B_0} \quad {\rm and} \quad W^{\mf{a}}_\phi \ket{B_0} \;.
$$
The cylinder amplitude for a closed string exchange between these D-branes is 
\be
\label{NonRatAmp0}
\bra{B_0}W^{\mf{g}}_{\mu^\ast} W^{\mf{a}}_{\sigma^\ast} W^{\mf{a}}_{\upsilon^\ast} q^{\frac{1}{2}(L_0 + \bar{L}_0 - \frac{c}{12})} W^{\mf{a}}_{\phi} \ket{B_0}  \;.
\ee
These boundary states are rational and have been described in \cite{Quella:2002ct}, where the Cardy condition was explicitly checked for the amplitude \eqref{NonRatAmp0}. It was shown that under the modular transform $q \rightarrow \tilde{q} = \exp -2\pi i\frac{1}{\tau}$, \eqref{NonRatAmp0} decomposes into a positive integer linear combination of the characters of $\hmf{a}_{xk} \otimes \hmf{g}_k/\hmf{a}_{xk}$.

But any D-brane on $G$ admits exactly marginal deformations. These marginal deformations can be seen as Goldstone modes: they arise from the global translational symmetries on the group which are broken by the insertion of the D-brane. They are implemented into our algebra $\mathcal{W}$ by the operators $t_g$. It seems very hard to imagine that an exactly marginal unitary perturbation of a consistent CFT could make it inconsistent. Therefore, after performing a marginal perturbation corresponding to a left translation of $W^{\mf{a}}_{\phi} B_0$ by $g$, we deduce that the amplitude  
\be
\label{NonRatAmp}
\bra{B_0}W^{\mf{g}}_{\mu^\ast} W^{\mf{a}}_{\sigma^\ast} W^{\mf{a}}_{\upsilon^\ast} q^{\frac{1}{2}(L_0 + \bar{L}_0 - \frac{c}{12})} t_g W^{\mf{a}}_{\phi} \ket{B_0}
\ee
should satisfy a non-rational version of Cardy's condition: after a modular transformation $q \rightarrow \tilde{q}$, \eqref{NonRatAmp} decomposes in a (possibly continuous) sum of character of a conformal vertex subalgebra of $\hmf{g}_k$ (see \cite{Gaberdiel:2001xm, Janik:2001hb}). Now, using the partition function identities reviewed above, we deduce that \eqref{NonRatAmp} is actually the partition function of open strings stretching between $W^{\mf{a}}_\upsilon t_g W^{\mf{a}}_{\phi} B_0$ and $W^{\mf{a}}_\sigma W^{\mf{g}}_\mu B_0$. This shows that the non-rational D-brane $W^{\mf{a}}_\upsilon t_g W^{\mf{a}}_{\phi} B_0$ should be consistent with all the D-branes preserving the rational subalgebra $\hmf{a}_{xk} \otimes \hmf{g}_k/\hmf{a}_{xk}$ of $\hmf{g}_k$. It is possible to iterate this argument and apply it to D-branes produced by the action of higher degree monomials in $\mathcal{W}$. Again, while this argument leaves little doubt about the existence of the non-rational D-branes generated by $\mathcal{W}$, the rigorous verification of the modular properties of $\eqref{NonRatAmp}$ seems out of reach.

\section{Geometric interpretation}

\label{SecGeom}

Now we will elucidate the geometry of the worldvolumes of the D-branes constructed in the previous section as boundary states. Such a geometrical interpretation is possible only in the limit $k \rightarrow \infty$, where the target space is the group manifold $G$ and D-branes can be pictured as submanifolds of $G$. At finite level, the algebra of functions on the target space of the quantum theory is finite dimensional and non-commutative, so no geometrical picture is available. We first briefly review some elementary notions of harmonic analysis on compact groups (see the chapters 3 and 4 of \cite{Hofmann1998Structure} for a precise and complete account).

\subsection{Harmonic analysis on compact groups}

Let $V$ be any finite dimensional irreducible $G$-module and let $\rho$ be the associated representation. Let $\phi \in {\rm End}(V) \sim V \otimes V^\ast$. We can associate to $\phi$ a complex-valued function $f_\phi$ on $G$ as follows:
\be
\label{HAEndFunc}
f_\phi(g) = {\rm Tr}_V(\phi \, \rho(g)) \;, \quad g \in G \;.
\ee
Conversely, any function $f$ on $G$ induces an endomorphism $\phi^V_f$ on $V$:
\be
\label{HAFuncEnd}
\phi^V_f(v) = \int_G dg f(g) \rho(g)v \;, \quad v \in V \;.
\ee
Let us write 
\be
\label{RefRGC}
R(G,\mathbbm{C}) = \bigoplus_{\lambda \in P^{\mf{g}}} V_\lambda \otimes V_{\lambda^\ast} \simeq \bigoplus_{\lambda \in P^{\mf{g}}} {\rm End}(V_\lambda)\;,
\ee
where $V_\lambda$ is the irreducible highest weight $G$-module of highest weight $\lambda$. \eqref{HAEndFunc} allows us to embed $R(G,\mathbbm{C})$ into the algebra $C(G,\mathbbm{C})$ of continuous complex-valued functions on $G$. The Peter-Weyl theorem states that the image of this embedding is dense in $C(G,\mathbbm{C})$. As a result, roughly speaking any continuous function $f$ on $G$ is determined by an operator $\phi_f$ on $\bigoplus_{\lambda \in P^\mf{g}} V_\lambda$ preserving the direct sum decomposition and vanishing on almost all $V_\lambda$. We can interpret this duality as a generalized Fourier transform.

The composition of elements of $R(G,\mathbbm{C})$ is related to the convolution product on $C(G,\mathbbm{C})$ by the Fourier transform. More precisely (cf. Lemma 3.43 in \cite{Hofmann1998Structure}):
\be
\label{ConvForm}
\phi^V_{f_1} \circ \phi^V_{f_2}(v) = \phi^V_{f_1 \ast f_2}(v) \;, \quad {\rm with} \; f_1 \ast f_2 (g) = \int_G dg' f_1(g') f_2(g'^{-1}g) 
\ee
This formula will be of crucial importance to us.

\subsection{The Fourier transforms of Wilson operators}

The restriction of any Wilson operator to the subspace of grade zero of $\mathcal{H}$ is an element of $R(G,\mathbbm{C})$, by the identification of the grade zero subspace of a $\hmf{g}_k$ module with a $\mf{g}$ module. It is therefore natural to ask what is its Fourier transform, as a function on $G$. We will answer this question in the semi-classical regime $k \rightarrow \infty$.

The Fourier transform of the delta distribution $\delta_g$ supported at a point $g \in G$ can be seen as an element in the completion $\bar{R}(G,\mathbbm{C})$ of $R(G,\mathbbm{C})$ obtained by dropping the finiteness condition implicit in the direct sum \eqref{RefRGC}. From \eqref{HAFuncEnd}, we have clearly
\be
\label{OpDeltaDist}
\phi_{\delta_g}|_{V_{\lambda}}(v) = \rho_\lambda(g) v \;, \quad v \in V_\lambda \;\; \mbox{for all} \; \lambda \in P^{\mf{g}} \;,
\ee  
where $\rho_\lambda$ is the representation of $G$ on $V_\lambda$. 

Let $C^A_{a_0}$ be the conjugacy class of $A \subset G$ passing through $a_0$. Consider the constant distribution 
$$
\delta_{C^A_{a_0}} = \frac{1}{{\rm Vol}(A)} \int_{A} da \, \delta_{a a_0 a^{-1}}
$$
supported on $C^A_{a_0}$. By linearity, the operator associated to it is 
$$
\phi_{C^A_{a_0}} = \frac{1}{{\rm Vol}(A)} \int_{A} da \,\phi_{\delta(a a_0 a^{-1})} = \frac{1}{{\rm Vol}(A)} \sum_{\lambda \in P^{\mf{g}}}  \int_{A} da \, \rho_\lambda(a a_0 a^{-1}) = 
$$
$$
= \frac{1}{{\rm Vol}(A)} \sum_{\lambda \in P^{\mf{g}}} \sum_{\sigma \in P^{\mf{a}}} \sum_{i = 1}^{b_{\lambda\sigma}} \int_{A} da \, \rho_{(\sigma,\lambda,i)}(a a_0 a^{-1}) \;,
$$ 
where we decomposed the irreducible $G$ representation $\rho_\lambda$ into a direct sum of irreducible $A$-representations $\rho_{(\sigma,\lambda,i)}$ of highest weight $\sigma$ and multiplicity $b_{\lambda\sigma}$. $\int_{A} da \, \rho_{(\sigma,\lambda,i)}(a a_0 a^{-1})$ commutes with the action of $A$, so must act by a multiple of the identity on the irreducible $A$-modules. Its eigenvalue is given by $\frac{1}{d_{\sigma}}{\rm Tr}_\sigma(a_0)$, where $d_\sigma$ is the dimension of $\rho_{(\sigma,\lambda,i)}$. Therefore:
$$
\phi_{C^A_{a_0}} = \frac{1}{d_{\sigma}}{\rm Tr}_\sigma(a_0) \quad {\rm on} \; V_{(\sigma,\lambda,i)} \subset V_\lambda \;.
$$

We would like to compare $\phi_{C^A_{a_0}}$ with a Wilson operator $W^{\mf{a}}_\tau$. We will make this comparison in a regime where $k \rightarrow \infty$, $\tau = k \tau_0$, with $\tau_0$ fixed. 
Note that if $k\tau_0$ is an integrable weight at level $k$, then $nk\tau_0$, $n \in \mathbbm{N}$, is an integrable weight at level $nk$, so the limit $k\rightarrow \infty$ makes sense. From \eqref{ActWLoop}, the eigenvalue of $W^{\mf{a}}_\tau$ on $V_{(\sigma,\lambda,i)}$ is given by 
$$
\chi_{\tau} \left ( -\frac{2\pi i}{k+\check{h}} (\sigma + \rho_{\mf{a}}) \right ) = \frac{S^{\mf{a}}_{\tau\sigma}}{S^{\mf{a}}_{0\sigma}} = \frac{S^{\mf{a}}_{\tau 0}}{S^{\mf{a}}_{00}} \frac{S^{\mf{a}}_{00}}{S^{\mf{a}}_{\sigma 0}} \frac{S^{\mf{a}}_{\sigma\tau}}{S^{\mf{a}}_{0\tau}} \;.
$$
Each factor on the right hand side can be expressed as a character, using the first equality. Taking the limit, we get:
$$
\lim_{k \rightarrow \infty} W^{\mf{a}}_{k\tau_0} = \frac{d_{\tau_0}}{d_\sigma} \chi_\sigma(-2\pi i\tau_0) \mathbbm{1} \quad {\rm on} \; V_{(\sigma,\lambda,i)} \;.
$$
So up to a global irrelevant factor $d_{\tau_0}$, $\lim_{k \rightarrow \infty} W^{\mf{a}}_{k\tau_0}$ coincides with $\phi_{C^A_{a_0}}$ if we identify $a_0 = (-2\pi i\tau_0)^\ast$. Therefore, the Fourier transform of $W^{\mf{a}}_{k\tau_0}$ is given in the semi-classical limit by a constant distribution supported on the conjugacy class $C^A_{(-2\pi i\tau_0)^\ast}$.

\subsection{The geometry of the non-rational WZW branes}

But now, using the convolution formula \eqref{ConvForm}, we can easily recover the geometry of any D-brane obtained from a known brane by the action of elements in the Wilson algebra. Suppose that we have a D-brane supported in the semi-classical limit on some submanifold $M \subset G$. Let $\delta_M$ be the Fourier transform of the grade zero component of the boundary state, a constant distribution supported on $M$. Acting by $t_{\tilde{g}}$ on the boundary state corresponds to a convolution of $\delta_M$ with $\delta_{\tilde{g}}$:
\be
\label{GeomActTg}
\delta_{\tilde{g}} \ast \delta_M = \int_G dg' \delta_{\tilde{g}}(g') \delta_M(g'^{-1}g) = \delta_M(\tilde{g}^{-1}g) = \delta_{\tilde{g}M}(g) \;,
\ee
so $t_{\tilde{g}}$, as expected, translates the worldvolume of the D-brane by the left action of $\tilde{g}$. Now suppose that we act with $W^{\mf{a}}_\sigma$ on the boundary state. The Fourier transform of the new boundary state is given up to a multiplicative constant by
\be
\label{GeomActWils}
\delta_{C^A_{a_0}} \ast \delta_M = \int_G dg' \delta_{C^A_{a_0}}(g') \delta_M(g'^{-1}g) = \delta_{M'}(g) \;,
\ee
where $M' = \{gg' | g \in C^A_{a_0}, g' \in M\}$. Therefore, geometrically, the action of $W^{\mf{a}}_\sigma$ amounts to change $M$ into the pointwise left multiplication of $M$ by the conjugacy class associated to $W^{\mf{a}}_\sigma$. This result fits with \cite{Quella:2002ct, Quella:2002ns}, where the worldvolumes of the D-branes \eqref{CosBoundSt} were shown to be products of conjugacy classes in the semi-classical limit. \\

Our geometric interpretation of defects is related to the one developed in \cite{Fuchs:2007fw}. There, the so-called folding trick \cite{1994NuPhB.417..403W} was used to see a defect as a D-brane on $G \times G$ (a ``bi-brane''), which is then pictured as a submanifold of $G \times G$ in the semi-classical limit. \cite{Fuchs:2007fw} associates to a maximally symmetric Wilson loop (with operator $W^{\mf{g}}_\mu$) the submanifold of $G \times G$ composed of all pairs $(g_1,g_2)$ such that $g_1g_2^{-1} \in C^G_{(-2\pi i\mu)^\ast}$, a ``biconjugacy class''. It turns out that this biconjugacy class is isomorphic to $C^G_{(-2\pi i\mu)^\ast} \times G$. On the other hand, the Fourier transform discussed above assigns directly the conjugacy class $C^G_{(-2\pi i\mu)^\ast}$ to $W^{\mf{g}}_\mu$.

A geometrical formula for the fusion of a defect with a D-brane was also conjectured in \cite{Fuchs:2007fw}. Suppose that the bi-brane associated with the defect has worldvolume $B \in G \times G$ and the D-brane has world volume $M \in G$, then their fusion should produce a D-brane with worldvolume $M'$, with 
$$
M' = p_{1}(p_{2}^{-1}(B) \cap M) \;.
$$
$p_i$ are the projection of the $i$th factor of $G \times G$ on $G$. Using the definition of the biconjugacy class, we see that this prescription reproduces \eqref{GeomActWils}. The convolution formula therefore provides a justification for this geometrical fusion formula, in the case of topological defects joining two copies of the same WZW theory in the semi-classical limit. \\

From an algebraic point of view, the restriction of Wilson operators to the grade zero subspace of $\mathcal{H}$, necessary to interpret them geometrically, is artificial. We are of course losing a lot of information in this process, and the convolution algebra of the corresponding functions/distributions is \emph{not} equivalent to the Wilson algebra. There exist a Peter-Weyl theorem for loop groups \cite{freed-2005, MR1646586}. However we do not know if an analog of the convolution formula holds in this case.

\section{Examples}

\label{SecExamples}

We already saw at the beginning of section \ref{SecNWZWB} that in the case $G = SU(2)$, the structure of the algebra is trivial and can be completely described. We now turn to the next examples in term of complexity, $SU(3)$ at level 1, $SU(3)$ at level 2 and $U\!Sp(4)$ at level 1.

Let us recall first (see section 14.2 of \cite{CFT1997}) that the outer automorphism group of $\hmf{su}_k(n)$ is the cyclic group of order $n$ acting the affine fundamental weights $\hat{\omega}_0 = (0,1)$, $\hat{\omega}_1 = (\omega_1,1)$, ... $\hat{\omega}_{n-1} = (\omega_{n-1},1)$. We denote between brace the projection of the affine weight on the weight space of $\mf{su}(n)$ and the level. As a result, if we use the usual parametrization of the level $k$ affine weights by the weights of $\mf{su}(n)$, the outer automorphism group permutes cyclically the weights $0, k\omega_1, ..., k\omega_{n-1}$. 

The eigenvalues of the Wilson operators displayed below are computed from \eqref{ActWLoop} and the explicit formulas for the modular S-matrix of affine Kac-Moody algebras (see for instance section 14.5 of \cite{CFT1997}.

\subsection{SU(3) at level 1}

\label{su3lev1}

\subsubsection{The structure of the algebra}

$\mf{su}(3)$ only has two types of nontrivial simple proper subalgebras, the regular $\mf{su}(2)$ subalgebras and the principal $\mf{su}(2)$ subalgebras. In the fundamental representation of $\mf{su}(3)$, they correspond to the embedding of a spin 1/2 or spin 1 representation of $\mf{su}(2)$. Their embedding index is respectively 1 and 4.

\paragraph{Maximally symmetric Wilson operators:}

In the case of $\hmf{su}_1(3)$, the only integrable weights are $0$, $\omega_1$ and $\omega_2$, and they are permuted by the outer automorphism group $\mathbb{Z}/3\mathbb{Z}$. As a result, the three maximally symmetric Wilson operators are given by
$$
W^{ms}_0 = \mathbbm{1}\;, \;\; W^{\rm ms}_{\omega_1} = t_z \;, \;\; W^{\rm ms}_{\omega_2} = t_{z^2} \;,
$$ 
where $z$ generates the center of $SU(3)$. Explicitly, they act by scalar multiplication on $H^{\hmf{su}_1(3)}_\nu$, with the following eigenvalues:
$$
\renewcommand{\arraystretch}{1.5}
\begin{array}{c|ccc}
\nu            & 0 & \omega_1 & \omega_2  \\
\hline 
W^{\rm ms}_0          & 1 & 1        & 1 \\
W^{\rm ms}_{\omega_1} & 1 & \kappa   & \kappa^2 \\
W^{\rm ms}_{\omega_2} & 1 & \kappa^2 & \kappa 
\end{array} \quad, \;\; \kappa = e^{2\pi i /3} \;.
$$

\paragraph{Regular $\mf{su}(2)$ Wilson operators:}

As the embedding index of a regular $\mf{su}(2)$ subalgebra is 1, the induced embedding of affine Kac-Moody algebras is $\hmf{su}_1(2) \subset \hmf{su}_1(3)$. $\hmf{su}_1(2)$ has only two integrable weights, 0 and $\omega$, and the outer automorphism group of $\hmf{su}_1(2)$ exchanges them. Therefore we have
$$
W^{\rm reg}_{0} = \mathbbm{1} \;,\;\; W^{\rm reg}_{\omega} = t_{z_{\rm reg}} \;,
$$
where $z_{\rm reg}$ is the image in $SU(3)$ of the nontrivial element of the center of $SU(2)$. The action of $W^{\rm reg}_{\omega}$ on $H^{\hmf{su}_1(2)}_\sigma$ is given by:
$$
\renewcommand{\arraystretch}{1.5}
\begin{array}{c|cc}
 \sigma  & 0 & \omega  \\
\hline W^{\rm reg}_{\omega} & 1 & -1
\end{array} \;\;.
$$

\paragraph{Principal $\mf{su}(2)$ Wilson operators:} 

The principal embedding of $\mf{su}(2)$ into $\mf{su}(3)$ induces an embedding $\hmf{su}_4(2) \subset \hmf{su}_1(3)$. This embedding turns out to be conformal: the $\hmf{su}_1(3)$ integrable modules $\hmf{su}_4(2)$ decompose into a finite number of $\hmf{su}_4(2)$ modules. More explicitly, the integrable highest weight modules of $\hmf{su}_4(2)$ are labelled by $n\omega$, $n = 0,1,2,3,4$, and we have:
\be
\label{DecompSU31SU24}
\renewcommand{\arraystretch}{1.8}
\begin{array}{c}
H^{\hmf{su}_1(3)}_0 \simeq H^{\hmf{su}_4(2)}_0 \oplus H^{\hmf{su}_4(2)}_{4\omega} \;,\\
H^{\hmf{su}_1(3)}_{\omega_1}, H^{\hmf{su}_1(3)}_{\omega_2} \simeq H^{\hmf{su}_4(2)}_{2\omega} \;.
\end{array}
\ee
The action of the principal $\mf{su}(2)$ Wilson operators $W^{\rm pr}_{\tau}$ on $H^{\hmf{su}_4(2)}_{\sigma}$ is given by:
\be
\label{ActWOPr}
\renewcommand{\arraystretch}{1.5}
\begin{array}{c|ccc}
\sigma               & 0        & 2\omega & 4\omega \\
\hline 
W^{\rm pr}_{\omega}  & \sqrt{3} & 0       & -\sqrt{3} \\
W^{\rm pr}_{2\omega} & 2        & -1      & 2 \\
W^{\rm pr}_{3\omega} & \sqrt{3} & 0       & -\sqrt{3} \\
W^{\rm pr}_{4\omega} & 1        & 1       & 1 
\end{array} \;\;.
\ee
We have 
$$
W^{\rm pr}_{4\omega} = t_{z_{\rm pr}} \;,
$$ 
where the $z_{\rm pr}$ is nontrivial element of the center of $SU(2)$ (which happens to act trivially on the $\hmf{su}_4(2)$ modules appearing in the decomposition \eqref{DecompSU31SU24}). From \eqref{DecompSU31SU24}, we see that $W^{\rm pr}_{2\omega}$ commutes with the whole action of $\hmf{su}_1(3)$. In fact,
$$
W^{\rm pr}_{2\omega} = W^{\rm ms}_{\omega_1} + W^{\rm ms}_{\omega_2} = t_z + t_{z^2} \;.
$$
Finally, we have from \eqref{ActWOPr}
$$
W^{\rm pr}_{\omega} = W^{\rm pr}_{3\omega} \;.
$$ 
(Note that this equality is true only due to the fact that the modules $H^{\hmf{su}_4(2)}_{\omega}$ and $H^{\hmf{su}_4(2)}_{3\omega}$ do not appear in the decomposition of the $\hmf{su}_1(3)$ modules.) \\

So far we managed to remove all the Wilson operators except $W^{\rm pr}_{\omega}$ from the presentation of the Wilson algebra. To understand the action of $W^{\rm pr}_{\omega}$, consider the outer automorphism generated by the symmetry of the Dynkin diagram of $\mf{su}(3)$. It extends to an automorphism\footnote{This automorphism does not appear in the outer automorphism group of $\hmf{su}_1(3)$, which is defined as the quotient of the symmetry group of the affine Dynkin diagram by the symmetry group of the finite Dynkin diagram. $\Omega$ generates the latter. This puzzling definition of the outer automorphism group of an affine Kac-Moody algebra allows to obtain an isomorphism with the center of $G$, see section 14.2 of \cite{CFT1997}.} $\Omega$ of $\hmf{su}_1(3)$. It is possible as well to define an action of $\Omega$ on the $\hmf{su}_1(3)$ integrable highest weight modules. The construction is straightforward: given a representation $\rho_\mu$ on a $\hmf{su}_1(3)$ module $H^{\hmf{su}(3)}_\mu$, define a new representation 
$$
\rho^\Omega := \rho_\mu \circ \Omega = \rho_{\Omega(\mu)} \;.
$$
The last equality implies that $H^{\hmf{su}(3)}_\mu$ endowed with the representation $\rho^\Omega$ is isomorphic as a $\hmf{su}(3)$ module to $H^{\hmf{su}(3)}_{\Omega(\mu)}$. If $\mu$ is a fixed point of $\Omega$, the latter induces a nontrivial endomorphism $\Omega_{\mu} : H^{\hmf{su}(3)}_\mu \rightarrow H^{\hmf{su}(3)}_\mu$. $\Omega_{\mu}$ acts trivially on the highest weight vector and satisfies the intertwining relation 
\be
\label{DefOmMu}
\Omega_{\mu}(Jv) = \Omega(J) \Omega_\mu(v) \;,
\ee
for $J \in \hmf{su}_1(3)$ and $v \in H^{\hmf{su}(3)}_\mu$. For more details about this construction, see for instance \cite{Fuchs:1995zr}. 

We claim that 
\be
\label{ActWOPr2}
W^{\rm pr}_{\omega} \simeq \sqrt{3} \Omega_0 \;\;{\rm on} \; H^{\hmf{su}_1(3)}_0 \;.
\ee
To check this claim, note that $\hmf{su}_4(2) \subset \hmf{su}_1(3)$ is the subalgebra left fixed by $\Omega$. $\Omega_0$ acts trivially on the highest weight vector $v_0$ of $H^{\hmf{su}_1(3)}_0$, so the submodule generated from $v_0$ by the action of $\hmf{su}_4(2)$ is an eigenspace of eigenvalue $+1$. 
On the other hand, if we denote by $J^{\alpha_1 + \alpha_2}_{-1}$ the generator of $\hmf{su}_1(3)$ of grade $-1$ associated with the sum of the two simple roots $\alpha_1$ and $\alpha_2$ of $\mf{su}(3)$, then $H^{\hmf{su}_4(2)}_{4\omega} \subset H^{\hmf{su}_1(3)}_0$ is generated from the vector $J^{\alpha_1 + \alpha_2}_{-1}v_0$. As $\Omega(J^{\alpha_1 + \alpha_2}_{-1}) = -J^{\alpha_1 + \alpha_2}_{-1}$, we deduce that $H^{\hmf{su}_4(2)}_{4\omega}$ is an eigenspace of $\Omega_0$ of eigenvalue $-1$. Comparing with \eqref{ActWOPr}, we get \eqref{ActWOPr2}.

From the definition \eqref{DefOmMu} of $\Omega_0$, we get the important relation 
\be
\label{CommutWg}
W^{\rm pr}_{\omega} t_g = t_{\Omega(g)} W^{\rm pr}_{\omega} \;.
\ee
Let us also remark that the square of $W^{\rm pr}_\omega$ is a linear combination of elements of the center:
\be
\label{SquareW}
(W^{\rm pr}_\omega)^2 = W^{\rm pr}_{2\omega} + \mathbbm{1} = t_z + t_{z^2} + \mathbbm{1} \;,
\ee
as can be easily computed from the fusion rules of $\hmf{su}_4(2)$.

We are now equipped to give a concrete description of the algebra $\mathcal{W}$ in the case $\hmf{g} = \hmf{su}_1(3)$. $\mathcal{W}$ has a family of generators $g \in SU(3)$ and an extra generator $W^{\rm pr}_{\omega}$. The relations \eqref{CommutWg} and \eqref{SquareW} allow us to write any element $W$ of $\mathcal{W}$ in the following form:
\be
\label{su31GenEl}
W = \sum_i n_i t_{g_i} + \sum_j n'_j t_{g'_j} W^{\rm pr}_{\omega} \qquad n_i, n'_j \in \mathbbm{N} \;, \;\; g_i, g'_j \in SU(3) \;.
\ee
We will now interpret this result in term of D-branes.

\subsubsection{Action on boundary states}

Let us act with the element $W$ defined in \eqref{su31GenEl} on a D0-brane state. Clearly, any term of the form $n t_g$ will create $n$ D0-branes\footnote{In the $\hmf{su}_1(n)$ theories, the only maximally symmetric Cardy type D-brane is the D0-brane.}, translated on the group manifold by the left action of $g$.

The terms of the form $n t_{g} W^{\rm pr}_{\omega}$ are more interesting. The boundary state corresponding to a D0-brane centered at the identity of the group manifold $SU(3)$ is given by
$$
\ket{B_0} = \kett{0}  + \kett{\omega_1} + \kett{\omega_2} \;.
$$
Using \eqref{ActWOPr} and the decomposition \eqref{DecompSU31SU24}, it is straightforward to compute the action of $W^{\rm pr}_{\omega}$ on $\ket{B_0}$ :
\be
\label{ActWMsToTw}
W^{\rm pr}_{\omega} \ket{B_0} = \left ( \frac{3}{2} \right )^{1/2}(\kett{0}_{\hmf{su}_4(2)} - \kett{4\omega}_{\hmf{su}_4(2)}) = 3^{1/4} \kett{0}^\Omega = \ket{B_0^\Omega} \;.
\ee
Using the conformal embedding \eqref{DecompSU31SU24}, we decomposed the Ishibashi state $\kett{0} = 2^{-1/2}(\kett{0}_{\hmf{su}_4(2)} + \kett{4\omega}_{\hmf{su}_4(2)})$ into Ishibashi states for $\hmf{su}_4(2)$, taking into account the normalization \eqref{NormIshi}. We then used the explicit action \eqref{ActWOPr} of $W^{\rm pr}_{\omega}$, which yields a multiple of the twisted Ishibashi state $\kett{0}^\Omega$. (The latter is normalized according to the conventions of \cite{Gaberdiel:2002qa}). \eqref{ActWMsToTw} is the boundary state of the D-brane twisted by the outer automorphism $\Omega$ \cite{Gaberdiel:2002qa}. Acting with a general monomial $n t_{g} W^{\rm pr}_{\omega}$ will produce $n$ copies of this twisted D-brane, translated on the group manifold by the left action of $g$. Therefore all the elements of the Wilson algebra produce either maximally symmetric or twisted D-branes when acting on the fundamental D0-brane state.

The reader familiar with D-brane charges on group manifolds might be puzzled by \eqref{ActWMsToTw}. From \eqref{FlowWLoop}, we deduce that there exists a Kondo flow $3\ket{B_0} \rightarrow W^{\rm pr}_{\omega} \ket{B_0} = \ket{B_0^\Omega}$ linking a stack of three maximally symmetric D0-branes to the twisted D-brane of $SU(3)$. But it is well known \cite{Maldacena:2001xj, Monnier:2008jj} that the twisted D-branes carry a different type of K-theory charges than the maximally symmetric D-branes, and therefore they should not be linked by any boundary renormalization group flow.

This paradox comes from the fact that we are considering a purely bosonic theory, whereas the boundary renormalization group flows admit conserved charges only in supersymmetric theories. Indeed, purely bosonic D-branes always have a tachyonic mode which allows them to decay into closed string radiation (see for instance \cite{Moore:2003vf}). By adding fermions transforming in the adjoint representation of the Lie group and performing a type 0 GSO projection, the WZW model can be turned into a supersymmetric theory, and all the known bosonic D-branes have worldsheet supersymmetric partners in the super WZW model. We refer the reader to the sections 3 and 4 of \cite{Monnier:2008jj} for a detailed account of these constructions.
In the supersymmetric theory, the D-brane obtained by the symmetry breaking Kondo flow described above and the twisted D-brane do not coincide anymore. Indeed, the closest supersymmetric analog of the bosonic $\hmf{su}_1(3)$ WZW model is the super WZW model based on $\hmf{su}_4(3)$, which admits as chiral algebra $\hmf{su}_1(3) \times \hmf{so}_1(8)$, the $\hmf{so}_1(8)$ part coming from the fermions. The twisted D-brane is the bosonic twisted D-brane \eqref{ActWMsToTw} tensored with a certain boundary state for $\hmf{so}_1(8)$. On the other hand, the fixed point of the symmetry breaking Kondo flow cannot be factorized in the same way. The construction of section 4.3 of \cite{Monnier:2008jj} shows that it preserves a subalgebra $\hmf{su}_{14}(2) \subset \hmf{su}_1(3) \times \hmf{so}_1(8)$, and clearly does not coincide with the twisted D-brane. \\

The next two examples will display nontrivial Wilson operators which are not associated to conformal embeddings, leading to a much larger Wilson algebra and new non-rational D-branes.

\subsection{SU(3) at level 2}

\paragraph{Maximally symmetric Wilson operators:}

The integrable weights of $\hmf{su}_2(3)$ are $0$, $\omega_1$, $\omega_2$, $2\omega_1$, $\omega_1 + \omega_2$, and $2\omega_2$. They form two orbits of the outer automorphism group: $\{0, 2\omega_1, 2\omega_2\}$ and $\{\omega_1, \omega_2, \omega_1 + \omega_2\}$. Therefore we have:
$$
W^{\rm ms}_0 = \mathbbm{1} \;, \;\; W^{\rm ms}_{2\omega_1} = t_z \;, \;\; W^{\rm ms}_{2\omega_2} = t_{z^2} \;, \;\; W^{\rm ms}_{\omega_1} = t_{z^2} W^{\rm ms}_{\omega_1 + \omega_2} \;, \;\; W^{\rm ms}_{\omega_2} = t_{z} W^{\rm ms}_{\omega_1 + \omega_2} \;,
$$
where again $z$ generates the center of $SU(3)$. There is a single independent maximally symmetric Wilson operator, which eigenvalues on the integrable weight modules $H^{\hmf{su}_2(3)}_\mu$ are given in the following table:
$$
\renewcommand{\arraystretch}{1.5}
\begin{array}{c|cccccc}
\mu & 0 & \omega_1 & \omega_2 & 2\omega_1 & \omega_1 + \omega_2 & 2\omega_2 \\
\hline W^{\rm ms}_{\omega_1 + \omega_2} & \phi & -\phi^{-1} & -\phi^{-1} & \phi & -\phi^{-1} & -\phi^{-1}
\end{array} \;\;,
$$
where $\phi = \frac{1}{2}(\sqrt{5} + 1)$ is the golden ratio.

\paragraph{Regular $\mf{su}(2)$ Wilson operators:}

We have a regular embedding $\hmf{su}_2(2) \subset \hmf{su}_2(3)$. The integrable weights of $\hmf{su}_2(2)$ are $0$, $\omega$ and $2\omega$, and the outer automorphism group exchanges $0$ and $2\omega$, so the only independent Wilson operator is $W^{\rm reg}_{\omega}$, which acts on $H^{\hmf{su}_2(2)}_\sigma$ with the following eigenvalues:
$$
\renewcommand{\arraystretch}{1.5}
\begin{array}{c|ccc}
\sigma & 0 & \omega & 2\omega  \\
\hline W^{\rm reg}_{\omega} & \sqrt{2} & 0 & -\sqrt{2}
\end{array} \;\;.
$$

\paragraph{Principal $\mf{su}(2)$ Wilson operators:}

Under the principal embedding, the $\hmf{su}_2(3)$ modules decompose into $\hmf{su}_8(2)$ modules with highest weight $0$, $2\omega$, $4\omega$, $6\omega$ and $8\omega$. The Wilson operators are labeled by the weights $n\omega$, $0 \leq n \leq 8$. The outer automorphism act as $n\omega \rightarrow (8-n)\omega$, leaving four independent Wilson operators. The action of $W^{\rm pr}_{\tau}$ on the modules $H^{\hmf{su}_8(2)}_\sigma$ is given by the following eigenvalues:
$$
\renewcommand{\arraystretch}{1.5}
\begin{array}{c|ccccc}
\sigma               & 0                 & 2\omega             & 4\omega & 6\omega             & 8\omega \\
\hline 
W^{\rm pr}_{\omega}  & \phi^{1/2}5^{1/4} & \phi^{-1/2}5^{1/4}  & 0       & -\phi^{-1/2}5^{1/4} & -\phi^{1/2}5^{1/4} \\
W^{\rm pr}_{2\omega} & \phi^2            & \phi^{-2}           & -1      & \phi^{-2}           & \phi^2   \\
W^{\rm pr}_{3\omega} & \phi^{3/2}5^{1/4} & -\phi^{-3/2}5^{1/4} & 0       & \phi^{-3/2}5^{1/4}  & -\phi^{3/2}5^{1/4} \\
W^{\rm pr}_{4\omega} & 2\phi             & -2\phi^{-1}         & 1       & -2\phi^{-1}         & 2\phi 
\end{array} \;\;.
$$
Using the fusion rules of $\hmf{su}_8(2)$, we have the relations:
\be
\label{RelPrWLSU23}
\renewcommand{\arraystretch}{1.5}
\begin{array}{c}
W^{\rm pr}_{2\omega} + \mathbbm{1} = (W^{\rm pr}_{\omega})^2  \;,\;\; W^{\rm pr}_{3\omega} + 2W^{\rm pr}_{\omega} = (W^{\rm pr}_{\omega})^3 \\
W^{\rm pr}_{4\omega} + 3(W^{\rm pr}_{\omega})^2 = (W^{\rm pr}_{\omega})^4  + \mathbbm{1} \;.
\end{array}
\ee
As we are working in an algebra over a semiring, we cannot perform the subtractions that would be needed to eliminate $W^{\rm pr}_{n\omega}$, $n = 2,3,4$ from the presentation. However, the equations \eqref{RelPrWLSU23} allow to translate any extra relation to be found for $W^{\rm pr}_{\omega}$ into relations for the other principal $\mf{su}(2)$ Wilson operators.
\\

In the case of $\hmf{su}_1(3)$, the conformal embedding allowed to compute the commutator of the nontrivial principal Wilson operator $W^{\rm pr}_\omega$ with the group defects $t_g$ \eqref{CommutWg}, what yielded a dramatic simplification of the Wilson algebra: it implied that any monomial can be expressed as a sum of monomials of degree 0 or 1 in $W^{\rm pr}_\omega$. No such relation occurs at level 2, and it is unclear to us whether other relations may hold. 

If our set is complete, a typical monomial in the Wilson algebra has the following form:
\be
\label{TypMonSU32}
n t_{g_{m+1}}W_m t_{g_m} W_{m-1}... W_2 t_{g_2} W_1 t_{g_1} (W^{\rm ms}_\mu)^p \;, 
\ee
where $m,n \in \mathbbm{N}$, $p \in \{0,1\}$, $g_i \in SU(3)$, $W_i = W^{\rm reg}_{\omega},\; W^{\rm pr}_{\tau}$. For this monomial to be irreducible, whenever $W_{i-1}$ and $W_i$ are associated to the same SU(2) subgroup, $g_i$ should not belong to the center of this subgroup.

Still assuming that no further relations hold, the action of these monomials on the D0-brane boundary state yields distinct boundary states. The D-brane produced by \eqref{TypMonSU32} is a stack of $n$ copies of the fundamental boundary state
$$
W_m t_{g_m} W_{m-1}... W_2 t_{g_2} W_1 (W^{\rm ms}_\mu)^p \ket{B_0} \;,
$$
translated on $SU(3)$ by the left action of $g_{m+1}$ and by the right action of $g_1$. The rational and previously known boundary states are those obtained from the monomials containing either a single or no symmetry breaking Wilson operator.

\subsection{USp(4) at level 1}

$\mf{sp}(4) \sim \mf{so}(5)$ has a two-dimensional weight space. The short simple root $\alpha_1$ of square norm 1 and the long simple root $\alpha_2$ of square norm 2 are at an angle of $3\pi/4$. The corresponding fundamental weights will be labeled $\omega_1$ and $\omega_2$. 

Clearly, the simple subalgebras of $\mf{sp}(4)$ can only be isomorphic to $\mf{su}(2)$. They are conjugate to one of the following subalgebras:
\begin{itemize}
	\item The ``long'' regular $\mf{su}(2)$ subalgebra which weight space is the line $\ell \alpha_2$, $\ell \in \mathbbm{R}$. It has an embedding index of 1. 
	\item The ``short'' regular $\mf{su}(2)$ subalgebra which weight space is the line $\ell \alpha_1$, $\ell \in \mathbbm{R}$. It has an embedding index of 2.
	\item The principal $\mf{su}(2)$ subalgebra which weight space is the line $\ell (3\alpha_1 + 2 \alpha_2)$, $\ell \in \mathbbm{R}$. It embedding index is 10, and it comes from the 5 dimensional real representation of $\mf{su}(2)$ in $\mf{so}(5)$.
\end{itemize}

\paragraph{Maximally symmetric Wilson operators:}

The integrable weights of $\hmf{sp}_1(4)$ are $0$, $\omega_1$ and $\omega_2$. The outer automorphism group exchanges $0$ and $\omega_1$, so $W^{\rm ms}_{\omega_2}$ acts like the nontrivial element of the center of $USp(4)$. The eigenvalues of $W^{\rm ms}_{\omega_1}$ on the modules $H^{\hmf{sp}_1(4)}_\mu$ are given by 
$$
\renewcommand{\arraystretch}{1.5}
\begin{array}{c|ccc}
\mu & 0 & \omega_1 & \omega_2  \\
\hline W^{\rm ms}_{\omega_1} & \sqrt{2} & 0 & -\sqrt{2}
\end{array} \;\;.
$$

\paragraph{Long regular $\mf{su}(2)$ Wilson operators:}

As the embedding index is 1, the corresponding Kac-Moody subalgebra is at level 1, and the Wilson operators can be identified with elements of the center of $SU(2)$. Therefore they can be eliminated from the presentation.

\paragraph{Short regular $\mf{su}(2)$ Wilson operators:}

The Wilson operators associated with the integrable weights $0$ and $2\omega$ of $\hmf{su}_2(2)$ are identified with the elements $1$ and $-1$ of the center of $SU(2)$. They are mapped by the embedding onto group elements of $USp(4)$ and can be eliminated from the presentation. $W^{\rm reg}_\omega$ acts as follows on $H^{\hmf{su}_2(2)}_\sigma$:
$$
\renewcommand{\arraystretch}{1.5}
\begin{array}{c|ccc}
\mu & 0 & \omega & 2\omega  \\
\hline W^{\rm reg}_{\omega} & \sqrt{2} & 0 & -\sqrt{2}
\end{array} \;\;.
$$
Using the fusion rules of $\hmf{su}_2(2)$, we have 
\be
\label{RelRegWLSp14}
(W^{\rm reg}_{\omega})^2 = 1 + t_{-1} \;.
\ee

\paragraph{Principal $\mf{su}(2)$ Wilson operators:}

The integrable weights of $\hmf{su}_{10}(2)$ are $n\omega$, $n = 0,...,10$, and the outer automorphism group exchanges $n\omega$ and $(10-n)\omega$. The embedding $\hmf{su}_{10}(2) \subset \hmf{sp}_1(4)$ is conformal and the branching rules are given by (cf. \cite{CFT1997}, section 17.5.2)
$$
\renewcommand{\arraystretch}{1.5}
\begin{array}{rl}
H^{\hmf{sp}_1(4)}_0 &\simeq H^{\hmf{su}_{10}(2)}_0 \oplus H^{\hmf{su}_{10}(2)}_{6\omega} \;,\\
H^{\hmf{sp}_1(4)}_{\omega_1} &\simeq H^{\hmf{su}_{10}(2)}_{3\omega} \oplus H^{\hmf{su}_{10}(2)}_{7\omega} \;, \\
H^{\hmf{sp}_1(4)}_{\omega_2} &\simeq H^{\hmf{su}_{10}(2)}_{4\omega} \oplus H^{\hmf{su}_{10}(2)}_{10\omega} \;.
\end{array} 
$$
The eigenvalues of the independent Wilson operators on $H^{\hmf{su}_{10}(2)}_\sigma$ are given by
$$
\renewcommand{\arraystretch}{1.5}
\begin{array}{c|cccccc}
\sigma               & 0             & 3\omega & 4\omega            & 6\omega            & 7\omega & 10\omega  \\
\hline 
W^{\rm pr}_{\omega}  & \psi          & 1       & \psi^{-1}          & -\psi^{-1}         & -1      & -\psi \\
W^{\rm pr}_{2\omega} & \sqrt{2} \psi & 0       & -\sqrt{2}\psi^{-1} & -\sqrt{2}\psi^{-1} & 0       & \sqrt{2} \psi \\
W^{\rm pr}_{3\omega} & \sqrt{3} \psi & -1      & -\sqrt{3}\psi^{-1} & \sqrt{3}\psi^{-1}  & 1       & -\sqrt{3} \psi \\
W^{\rm pr}_{4\omega} & \psi^2        & -1      & \psi^{-2}          & \psi^{-2}          & -1      & \psi^2 \\
W^{\rm pr}_{5\omega} & 2\psi         & 0       & 2\psi^{-1}         & -2\psi^{-1}        & 0       & -2\psi
\end{array} \;\;,
$$
where $\psi = \frac{\sqrt{3}+1}{\sqrt{2}}$. Like in the case of $\hmf{su}_2(3)$, the fusion rules of $\hmf{su}_{10}(2)$ allow to deduce relations between the operators $W^{\rm pr}_{\sigma}$, so that practically we only have to understand the action of $W^{\rm pr}_{\omega}$. They read:
\be
\label{RelPrWLSp14}
\renewcommand{\arraystretch}{1.5}
\begin{array}{c}
W^{\rm pr}_{2\omega} + \mathbbm{1} = (W^{\rm pr}_{\omega})^2  \;,\;\; W^{\rm pr}_{3\omega} + 2W^{\rm pr}_{\omega} = (W^{\rm pr}_{\omega})^3 \;,\\
W^{\rm pr}_{4\omega} + 3(W^{\rm pr}_{\omega})^2 = (W^{\rm pr}_{\omega})^4  + \mathbbm{1} \;.
\end{array}
\ee
More interestingly, the conformal embedding allows us to deduce the relations 
\be
\label{RelPrWLSp14b}
W^{\rm pr}_{2\omega} = W^{\rm ms}_{\omega_1} W^{\rm pr}_{\omega} \quad {\rm and} \quad W^{\rm pr}_{5\omega} = (W^{\rm ms}_{\omega_1})^2 W^{\rm pr}_{\omega} \;.
\ee
Combining the first equations of \eqref{RelPrWLSp14} and \eqref{RelPrWLSp14b}, we get 
\be
\label{RelPrWLSp14c}
(W^{\rm pr}_{\omega})^2 = W^{\rm ms}_{\omega_1} W^{\rm pr}_{\omega} + \mathbbm{1} \;,
\ee
so that any power of $W^{\rm pr}_{\omega}$ can be reduced to terms of degree zero or one in $W^{\rm pr}_{\omega}$. \\
\\
With the relations derived so far, we have reduced the generators of the Wilson algebra of $\hmf{sp}_1(4)$ to the set
$$
\{ t_g , \; W^{\rm ms}_{\omega_1} , \; W^{\rm reg}_{\omega} , \; W^{\rm pr}_{\sigma} \} \;,\;\; g \in U\!Sp(4) \;, \;\; \sigma = \omega, \;...\,, \; 5\omega \;.
$$
With the relations \eqref{RelRegWLSp14}, \eqref{RelPrWLSp14}, \eqref{RelPrWLSp14c} in addition to the universal relations in \eqref{GenRel}. The situation is similar to the one encountered for $SU(3)$ at level 2: as $U \!Sp(4)$ does not have any outer automorphism and as $W^{\rm reg}_{\omega}$ and $W^{\rm pr}_{\sigma}$ clearly do not coincide with inner automorphisms, no relation of the form \eqref{CommutWg} can hold for all $g \in U\!Sp(4)$. However we cannot prove that the relations we found form a complete set.

If no further relations hold, a generic monomial in the algebra involves products of an arbitrary large number of generators, of the form
$$
n \, t_{g_{m+1}} W_m t_{g_m}\,...\, W_2 t_{g_2} W_1 t_{g_1} (W^{\rm ms}_{\omega_1})^p \;,
$$
where $n,m \in \mathbbm{N}$ , $p \in \{0,1\}$, $g_i \in U\!Sp(4)$ and $W_i =W^{\rm reg}_{\omega}, \, W^{\rm pr}_{\sigma}$. $g_i$ does not belong to the center of $U\!Sp(4)$ if $W_i$ and $W_{i+1}$ are associated to the same subalgebra of $\mf{sp}(4)$.

\section{Generalizations}

\label{SecGener}

Finally, we briefly discuss generalizations of our results to WZW models with non-diagonal modular invariants and coset models.

So far we considered only WZW models with charge conjugation modular invariant, whose state space is given by \eqref{StSpChCModIn}. There exist models with different state spaces, whose torus partition function is nevertheless modular invariant. $\mathcal{W}$ is purely chiral, so it acts on the state space of WZW models with any modular invariant, as well as on the boundary states of these models. It may happen that some integrable modules are missing in the state space of the models with nontrivial modular invariants. In this case, additional relations between elements of $\mathcal{W}$ may appear, because the Wilson algebra was not defined intrinsically, but rather through the action of the Wilson operators on the state space. A phenomenon of this type was encountered for the Wilson loops associated to the principal $\mf{su}(2)$ subalgebras of $\mf{su}(3)$, see the table \eqref{ActWOPr}. Moreover, one may want to consider as well the ``antiholomorphic'' Wilson algebra, defined with the Wilson operators and group defects built out of the antiholomorphic current $\bar{J}$. Indeed, its action does not necessarily coincide with the action of the holomorphic Wilson algebra for models with nontrivial modular invariant. Still, the two actions always commute, so no new difficulty should appear in the analysis. 
\\
\\
Coset models can be seen as WZW models for which a subgroup $H \subset G$ has been gauged. Consistency with the gauging requires the boundary states and defect operators to be invariant under the action of the affine Kac-Moody subalgebra $\hmf{h}_{yk} \subset \hmf{g}_k$, where $y$ is the embedding index of $\mf{h}={\rm Lie}(H) \subset \mf{g}$. The state space of the gauged WZW model is built out of the Hilbert spaces $H^{\hmf{h}_{yk}/\hmf{g}_k}_{(\mu,\sigma)}$ defined under \eqref{DecompStSpace}. Only the $\hmf{h}_{yk}$-invariant part $\mathcal{W}^{\hmf{h}_{yk}}$ of $\mathcal{W}$ can act on $H^{\hmf{h}_{yk}/\hmf{g}_k}_{(\mu,\sigma)}$. It is not difficult to find a set of generators for $\mathcal{W}^{\hmf{h}_{yk}}$. They are given by:
\begin{itemize}
	\item $t_g$, with $g$ belonging to the centralizer of $H$ in $G$.
	\item $W^{\mf{a}}$, with $\mf{h} \subseteq \mf{a} \subseteq \mf{h}$.
	\item $W^{\mf{a}}$ with $[\mf{h}, \mf{a}] = 0$.
\end{itemize}
The relations of \eqref{GenRel} still hold. This construction will not provide any new D-brane for coset models associated with maximal Lie group embeddings.

\subsection*{Acknowledgements}

I would like to thank Anton Alekseev, Stefan Fredenhagen, Thomas Quella and Ingo Runkel for discussions. Special thanks go to Thomas Quella for detailed comments on a draft of this paper. This project is supported by the Swiss National Science Foundation, grant PBGE2--121187.

\appendix

\section{Algebras over a commutative semiring}

The definition of an algebra over a commutative semiring, though elementary, is not easily found in the litterature. We recall it here. 
\begin{definition}
\textit{A commutative semiring $R$ is a set endowed with two operations, addition $(+)$ and multiplication $(*)$, such that
\begin{itemize}
	\item $(R,+)$ is a commutative monoid with $0_R$ as identity element;
	\item $(R,*)$ is a commutative monoid with identity element $1_R$;
	\item the multiplication is distributive over the addition from either side;
	\item $0_R * a = a * 0_R = 0_R$ for all $a \in R$;
	\item $0_R \neq 1_R$.
\end{itemize}}
\end{definition}
The fact that the additive law provides $R$ with a structure of monoid instead of a group structure implies that the addition is not necessarily invertible. The basic example of a commutative semiring is given by the natural integers $\mathbbm{N}$, endowed with the familiar addition and multiplication.

\begin{definition}
\textit{A semimodule over a commutative semiring $R$ is a commutative monoid $(M,+)$ with identity element $0_M$ endowed with a scalar multiplication $R \times M \rightarrow M$, $(r,m) \rightarrow rm$ such that
\begin{itemize}
	\item $(rr')m = r(r'm)$;
	\item $r(m+m') = rm + rm'$;
	\item $(r+r')m = rm + r'm$;
	\item $1_Rm = m$;
	\item $r0_M = 0_M = 0_Rm$
\end{itemize}
for all $r,r' \in R$, $m,m' \in M$.}
\end{definition}
The set of D-branes forms a semimodule over $\mathbbm{N}$, where $n \in \mathbbm{N}$ acts on a D-brane $B$ by producing a stack of $n$ copies of $B$.

\begin{definition}
\textit{An algebra $A$ over a commutative semiring $R$ is a semimodule over $R$ endowed with a bilinear product $\ast$:
$$
a\ast(a' + a'') = a\ast a'+ a\ast a'' \;, \quad (a'+a'')\ast a = a' \ast a + a'' \ast a
$$
$$
(ra) \ast a' = r(a \ast a') = a \ast (ra') \;,
$$
for all $a,a',a'' \in A$ and $r \in R$. If the product is associative or admits a unit, $A$ is said to be associative or unital.}
\end{definition} 
The set of topological defects form an associative unital algebra over $\mathbbm{N}$. One can define a (not necessarily commutative) semiring by dropping the commutativity condition on the multiplicative law, and check that an associative unital algebra over a commutative semiring is in fact a semiring. In this paper, we prefer to see the topological defects as an algebra over $\mathbbm{N}$ to emphasize the natural scalar multiplication by positive integers, corresponding to stacking identical defects.

\label{SecSemi}

{
\small
\providecommand{\href}[2]{#2}\begingroup\raggedright\endgroup
}

\end{document}